\documentclass[pra,floatfix,showpacs,amsmath,twocolumn]{revtex4-1}

\makeatletter
\let\@ORGREVTEXendnotemark\@endnotemark
\let\@ORGREVTEX@makefnmark@cite\@makefnmark@cite
\def\@endnotemark{\bgroup\@fileswfalse\@ORGREVTEXendnotemark\egroup}
\def\@makefnmark@cite{\bgroup\@fileswfalse\@ORGREVTEX@makefnmark@cite\egroup}
\makeatother

\usepackage{amssymb}
\usepackage[mathscr]{eucal}
\usepackage{graphicx}

\usepackage{braket}
\usepackage{enumerate}
\usepackage{subfigure}
\usepackage{multirow}

\usepackage{mciteplus}
\mciteErrorOnUnknownfalse

\usepackage{color}

\def\tr{\mbox{Tr}}

\begin{document}

\title{Matrix product states for critical spin chains: \\
finite size scaling versus finite entanglement scaling}

\author{B. Pirvu$^1$,  G. Vidal$^3$, F. Verstraete$^1$, L. Tagliacozzo$^2$}
\affiliation{$^1$Fakult\"at f\"ur Physik, Universit\"at Wien,
Boltzmanngasse 5, A-1090 Wien, Austria\\
$^2$ICFO, Insitut de Ciencias Fotonicas, 08860 Castelldefels (Barcelona),
Spain\\
$^3$ Perimeter Institute for Theoretical Physics, Waterloo, Ontario, N2L2Y5, Canada}

\date{\today}

\begin{abstract}
We investigate the use of matrix product states (MPS) to approximate
ground states of critical quantum spin chains with periodic boundary
conditions (PBC). We identify two regimes in the $(N,D)$ parameter plane,
where $N$ is the size of the spin chain and $D$ is the dimension of the
MPS matrices. In the first regime MPS can be used to perform finite size scaling (FSS). In the complementary regime the MPS simulations show
instead the clear signature of finite entanglement scaling (FES).
In the thermodynamic limit (or large $N$ limit), only MPS in the
FSS regime maintain a finite overlap with the exact ground state. This observation has implications on how to correctly perform FSS with MPS, as
well as on the performance of recent MPS algorithms for systems with PBC.
It also gives clear evidence that critical models can actually be
simulated very well with MPS by using the right scaling relations; in
the appendix, we give an alternative derivation of the result of
Pollmann et al. [Phys. Rev. Lett. 102, 255701 (2009)] relating the
bond dimension of the MPS to an effective correlation length.
\end{abstract}

\maketitle


\section{Introduction}
\label{sec:introduction}

Quantum many body systems are very hard to study due to the exponential
growth of their Hilbert space with the number of constituents. One
possible cure to this issue for one dimensional systems is to describe
their ground states as matrix product states (MPS)
\cite{kluemper-1993,frank-2008-review,schollwoeck-2011-review}.
This family of states is known to be well suited to study gaped 1D
phases~\cite{fannes-1992} where for generic systems almost exact results
can be obtained with matrices whose size does not depend on the size of
the system. Even more, for several gapped 1D systems the exact
ground state can be expressed in terms of translationally invariant
MPS with very small bond dimension~\cite{aklt-1987,fannes-1992}.
Gapless 1D phases are harder to simulate with MPS since the size of the
matrices necessary to obtain good approximations of their ground states
increases polynomially with the size of the system.
This is particularly unfortunate since the universal low energy information
encoded in the gapless phase becomes apparent only for large systems.

Luckily such universal information is also encoded in the way a state
approaches the thermodynamic limit and one can extract it by using
the celebrated finite size scaling (FSS) technique~\cite{book-barber-1983}.
This technique amounts to study larger and larger systems in a gapless
phase (that due to the finite size of the system becomes gapped) and
extract universal properties through the dependence of the observables
on the system size.

In the context of MPS, one can use an alternative approach to study
gapless phases. It is called finite entanglement scaling (FES)
\cite{tagliacozzo-2008} and amounts to study the scaling of the
expectation value of observables in the ground state of infinite
chains described by MPS with fixed bond dimension and thus
finite entanglement
\footnote{
Additionally, one can also deform the system by describing it on curved
geometries where the curvature induces a gap, and then study the approach
to the flat limit in the same way one would study the approach
to an infinite system \cite{ueda_hyperbolic_2009}.
}.

Both the existence of FSS and FES close to a conformal fixed point are a
direct consequence of conformal invariance
\cite{bloete-cardy-nightingale-1986,moore-2008}.
If $N$ is the chain length and $D$ the MPS bond dimension, then FSS
corresponds to taking $D\to\infty $ first and then taking $N\to\infty$,
whereas FES consists in taking $N\to\infty$ first and then $D\to\infty$.

An important question to ask is whether FSS and FES provide the same
universal information. Since the proposal of FES for simulations with MPS
\cite{tagliacozzo-2008} it has been shown that indeed quantities such
as critical exponents related to local observables or the central charge
of the model can be extracted with the help of this technique
\cite{zheng_continuous_2011,rodriguez_field-induced_2011,
dai_ground-state_2010,orus_geometric_2010,mcculloch_infinite_2008,
heidrich-meisner_phase_2009,zhao_spontaneous_2010}, in a
similar way as it is normally done with FSS techniques.
Here we will show, however, that some care is required in order to
differentiate between the effects of FES and those of FSS.

Specifically, we consider critical systems with periodic boundary
conditions (PBC), and describe their translationally invariant ground
states using translationally invariant MPS. In order to perform FSS one
should obtain for each system size $N$ a sequence of increasingly accurate
MPS approximations with growing bond dimension $D$, which for large enough
$D$ converges to the exact ground state. Importantly, we find that for an
intermediate range of values of $D$, for which local observables are
already reproduced with high accuracy and show scaling behavior, the MPS
approximation is almost completely orthogonal to the exact ground state
(resulting e.g. in failure to reproduce correlation functions at distance
$N/2$, as previously illustrated in the inset of Fig. 11 and 12 of Ref.
\cite{me-2010-PBCI}). In other words, reasonably converged values of
(and/or scaling behavior for) local observables including the ground state
energy, are not sufficient criteria to establish that some MPS is a good
approximation to the ground state of a critical PBC system. Instead, in
order to properly apply FSS, for each system size $N$ one should consider
MPS with a bond dimension $D$ larger than some threshold value $D_0$,
where $D_0$ depends both on $N$ and on the spin model.

Our results have important consequences for the design of algorithms
that simulate PBC chains with MPS. Simulating PBC systems with MPS is
computationally much more expensive~\cite{frank-2004a} than simulating
the same system with open boundary conditions (OBC). Nonetheless, when
studying critical ground states, systems with PBC are known to approach
the thermodynamic limit much faster than systems with OBC, and therefore
they offer a much better framework for FSS.
For this reason, substantial effort \cite{pippan-2010,zhou-2009,me-2010-PBCI,rossini-2011} has been made to
try to lower the computational cost of MPS simulations
with PBC. Two types of approaches have been pursued.
One consists in building a MPS for a finite system with PBC by using the
translationally invariant MPS tensor that has been optimized in an
infinite chain with OBC~\cite{zhou-2009}.
This approach is equivalent to a crude approximation of the MPS
transfer matrix: the $D^2 \times D^2$ matrix is approximated only by its
dominant eigenvector
\footnote{
Eigenvector corresponding to the eigenvalue with the largest magnitude.
}.
The second approach~\cite{pippan-2010,me-2010-PBCI,rossini-2011} accounts
for PBC by retaining more than one eigenvector in the approximation of
the transfer matrix.
We show in this work that the first approach fails to provide an
accurate ground state approximation for critical PBC systems.
A detailed comparison of these algorithms can be found in
Appendix~\ref{sec:other_algorithms}.

We will build our arguments by studying two paradigmatic critical spin
chains: the quantum Ising model (IS) and the quantum Heisenberg (HB) model,
for chains with PBC and linear size $N$. The ground states are encoded in
MPS of a given bond dimension $D$. Even if the Hamiltonian is critical,
both the finite size of the chain and the finite bond dimension of the MPS
induce a gap

\begin{equation}\label{eq:fsg}
  \Delta E_N = \xi_{N}^{-1}=\frac{2\pi \,x_{1}}{N}
\end{equation}

\begin{equation}\label{eq:feg}
  \Delta E_D = \xi_{D}^{-1}\propto D^{-\kappa}
\end{equation}
\noindent
where $x_{1}$ is the smallest critical exponent of the theory
\cite{cardy_operator_1986} and $\kappa$ is the exponent for the scaling
of the effective correlation length of MPS simulations with finite
bond dimension~\cite{moore-2008,tagliacozzo-2008}.
Depending on which of the two gaps dominates, the system is in
one of the two regimes

\begin{equation}\label{eq:fsr}
  \xi_{D}\gg\xi_{N}: \textrm{FSS regime}
\end{equation}

\begin{equation}\label{eq:fer}
  \xi_{N}\gg\xi_{D}:\textrm{FES regime}
\end{equation}
\noindent

The presence of  two regimes in the PBC chain can be intuitively
understood in the following way: in the FES regime defined by
equation (\ref{eq:fer}) the small dimension of the MPS matrices
implies that the system is not aware of its geometry. Thus the boundaries
do not play any role. In the FSS regime, defined by
equation (\ref{eq:fsr}) on the other hand, the size of the matrices
is big enough to notice the presence of the boundaries and thus
different choices of boundary conditions lead to different MPS.

%
%

For simulations where

\begin{equation}\label{eq:transition}
  \xi_{N}\simeq\xi_{D}
\end{equation}
\noindent
we find for all values of $N$ and $D$ that are accessible numerically
the presence of an abrupt transition between the FSS and FES regimes
(for a related work see also Ref.~\cite{ling-sandvik-2010}).
One way to observe this transition is by looking at the difference between
the exact ground state energy in the thermodynamic limit
and the energy of MPS approximations with different $N$ and $D$.
For fixed $D$ these plots show a steep transition between the FSS regime
where the difference scales like $\propto N^{-2}$ to the FES regime
where the difference does not depend on $N$.
Another way is to look at the overlap between MPS with different
$D$ for fixed chain length $N$:
starting off with a MPS with some big $D_{max}$, we look at its overlap
with MPS with decreasing $D$. We then observe how the initially
smoothly decreasing overlap abruptly drops towards lower values at some
$D_r$, unambiguously showing the transition to the FES regime.
Now the overlap is a global variable and as such indeed aware of the
boundary conditions. The main finding
of this paper is that states in the FES regime, while possessing the same
local universal properties~\cite{tagliacozzo-2008} like those in the FSS
regime, turn out to have vanishing overlap with them.

We also present a possible technique to determine if a given bond
dimension is sufficient to enter the FSS regime, so that we can give
the computationally most favorable recipe to access global universal
properties that depend on the boundaries
(for a discussion of these properties see e.g.
\cite{evenbly-pfeifer-luca-guifre-2010}).

\begin{figure*}[ht]
  \begin{center}
    \includegraphics[width=1.0\textwidth]{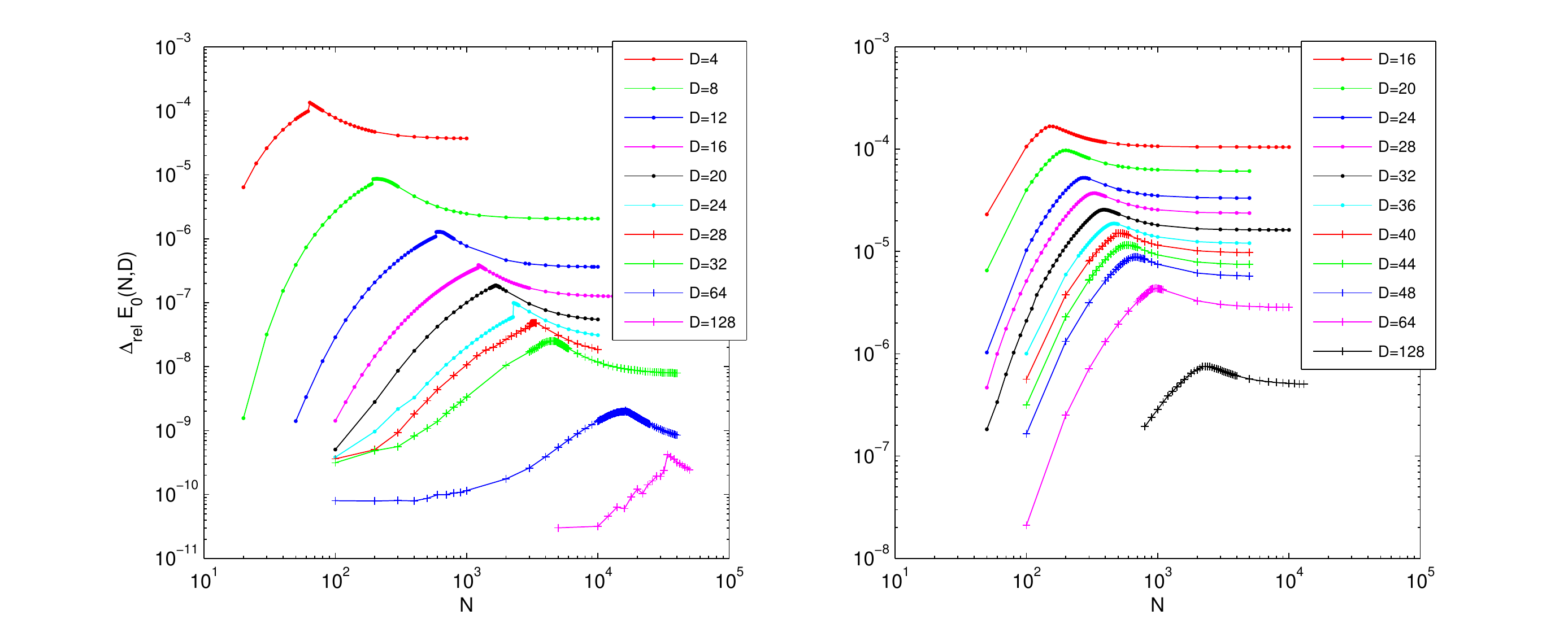}
  \end{center}
  \caption{
    (Color online).
    Quantum Ising (left) and Heisenberg model (right):
    Relative precision of the optimal MPS ground state energy for
    different $D$ as a function of $N$. The position of the \emph{hump}
    that can be observed between the small and big $N$ limits of each
    curve turns out to be proportional to the effective correlation
    length $\xi_D$ of MPS with finite bond dimension.
  }
\label{fig:ISHB_Nall_Dall}
\end{figure*}

The paper is organized as follows. We start by introducing the IS and HB
models as well as the technique used to simulate them in section
\ref{sec:overview}. In section \ref{sec:two_regimes} we present numerical
evidence for the presence of the FSS and FES regimes in MPS simulations
of PBC chains by looking at the ground state energy.
In section \ref{sec:transition} we discuss how to identify the sharp
transition between the two regimes by looking at the overlap.
Then, in section~\ref{sec:correction_to_scaling}, we give a more detailed
view on the transition between the two scaling regimes.
Section \ref{sec:minimal_D_faithful_simulations} gives a recipe for
obtaining the minimal bond dimension needed to observe global
universal properties of critical systems.
In section \ref{sec:IS_TL} we perform a numerical study of the transition
and for the IS model we can provide evidence for its persistence in the
thermodynamic limit. For the HB model we are not able to do the same
due to the coarser precision of our simulations for this model.
In section \ref{sec:scaling_function} we provide a numerical analysis
of the scaling function for the energy difference that reveals a
two-parameter scaling analogous to the one found in the context
of critical 2D classical systems by Nishino in Ref.~\cite{nishino-1996}.
We conclude with a discussion of the implications of our results and with
a brief outline of future developments.

All technical details are contained in the appendices. There we first
provide an alternative way to derive the analytical result for the scaling
exponent $\kappa$ in $\xi_D\propto D^\kappa$,
which we find more intuitive than the one given in~\cite{moore-2008}.
Then we show how our algorithm can be used in order to extract $\kappa$
from the numerical results for the ground state energy.
For the IS model, we are able to provide a numerical
confirmation for the persistence of the transition between the FSS
and the FES regime in the thermodynamical limit.

\section{Numerical results}
\label{sec:overview}
We will use throughout this work the algorithm presented
in~\cite{me-2010-PBCI}. That algorithm exploits the translational
invariance of the models we study by using an ansatz
based on translationally invariant MPS. This means that the MPS
tensors at each site of the chain are identical thus reducing  the cost of the simulation by a factor $N$. The energy is
minimized by means of a conjugate gradient method in the subspace
spanned by real and symmetric MPS with bond dimension $D$.
The computational cost scales like $O(mnD^3)+O(n^2 D^3)$.
$m$ and $n$ are parameters whose magnitude depends on the
entanglement of the ground state of the model under consideration.
For more details on the method we refer the reader to that work.
The two paradigmatic models we have considered are the critical quantum
Ising model described by the Hamiltonian

\begin{equation}\label{eq:H_IS_sim}
  H_{IS} = - \sum_{i=1}^{N} \sigma_i^z \sigma_{i+1}^z
           - \sum_{i=1}^{N} \sigma_i^x
  \,\,\,,
\end{equation}
\noindent
and the Heisenberg model described by the Hamiltonian

\begin{equation}\label{eq:H_HB}
  H_{HB} = \sum_{i=1}^{N}  \vec{S}_i \vec{S}_{i+1}
  = \frac{1}{4} \sum_{i=1}^{N}
   (\sigma^x_i \sigma^x_{i+1} + \sigma^y_i \sigma^y_{i+1}
   +\sigma^z_i \sigma^z_{i+1})
\end{equation}
\noindent
where the 1D lattice is considered to be periodic. Both Hamiltonians are critical which means that their gap between the ground state and the first excited state closes as an inverse power of $N$ as described in Eq. \ref{eq:fsg}.

As a matter of fact the analysis in this work was triggered by
a comprehensive study of the precision of the algorithm presented in
\cite{me-2010-PBCI}. We originally wanted to assess the usability
of that method and to this end we simulated a plethora of different
configurations $\{N,D\}$ for the IS and the HB models.
A selection of these simulation results is shown in
Figure~\ref{fig:ISHB_Nall_Dall} where we plot the relative energy
precision of simulations with different $D$ for many different
chain lengths $N$.

The shape of curves with constant $D$ is very surprising since it shows
a fundamental deviation from what we would have expected. Our expectation
was that for short chains the precision will be generally better than
for long chains and that as $N$ gets bigger and bigger, the precision
will eventually saturate from below to the value obtained with the
corresponding $D$ when simulating the chain in the thermodynamic limit.
Obviously the small $N$ and the big $N$ regimes are in accordance with our
expectation. However at some point between these limits we see the
emergence of a \emph{hump} which indicates that something interesting
is happening in that region. As a matter of fact we can show that if
we interpret the position of the \emph{hump} as an indicator for
the effective correlation length $\xi_D$ of MPS
with finite $D$, we can reproduce the theoretically
predicted result for the scaling of $\xi_D$~\cite{moore-2008} with
very good accuracy (see Appendix~\ref{sec:eff_corr_len}).
The results  presented in this work provide the general framework to  understand the emergence of the \emph{hump}
and to explain what is happening when moving from the left
to the right side of Fig. ~\ref{fig:ISHB_Nall_Dall}.

\subsection{Two different regimes for MPS simulations}
\label{sec:two_regimes}

As already mentioned in the introduction a MPS simulation close to the
critical point is an example of a two scale problem. This is not something
unexpected as it has been already pointed out in the context of
2D classical systems studied with the corner transfer matrix by Nishino and
coworkers~\cite{nishino-1996} and in the context of quantum phase
transitions in 1D quantum chains with open boundaries by one of the
authors (section IIIG of Ref.~\cite{tagliacozzo-2008}).
In the scenario we are considering, the two scales appearing are i) the
correlation length induced by the finite size of the system $N$ of Eq.
\ref{eq:fsg} and ii) the correlation length induced by the size of the
matrices $D$ of Eq. \ref{eq:feg}. Depending on the relation among the two
stated in Eqs. \ref{eq:fsr} and \ref{eq:fer}, the system will be in one of
the two different regimes, respectively the FSS regime or the FES regime.

The approach to the thermodynamic limit of the ground state energy, as
function of the relevant parameter $N$ or $D$, is very different in the
two regimes so that we can use it as a footprint for them.
In the FSS regime indeed it obeys the celebrated result by Cardy and
Affleck~\cite{bloete-cardy-nightingale-1986, affleck-1986} from
conformal field theory (CFT),

\begin{equation}\label{eq:CFT_Escaling_PBC}
  E_0(N)-E_0(\infty)=-\frac{v_f \pi c}{6 N^2}
  \,\,\,,
\end{equation}
\noindent
where $E_0(\infty)$ is the thermodynamic limit, $v_f$ is the Fermi
velocity and $c$ is the central charge of the considered model.
In the thermodynamic limit, several authors have reported that
\cite{tagliacozzo-2008,me-2010-MPOR,moore-2008}

\begin{equation}\label{eq:MPS_Escaling_TL}
  E_0(D)-E_0(\infty) \propto \frac{\Delta}{D^\omega}
  \,\,\,.
\end{equation}
\noindent
where $\omega = 2\kappa$ and $\kappa$ is the same exponent of
Eq. \ref{eq:feg} and $\Delta$ is a positive non-universal constant.
We show below that the same scaling holds also
for MPS simulations of finite chains if $N$ is big enough. This happens
exactly in the FES regime defined by (\ref{eq:fer}).

\begin{figure}[ht]
  a)\includegraphics[width=1.0\columnwidth]{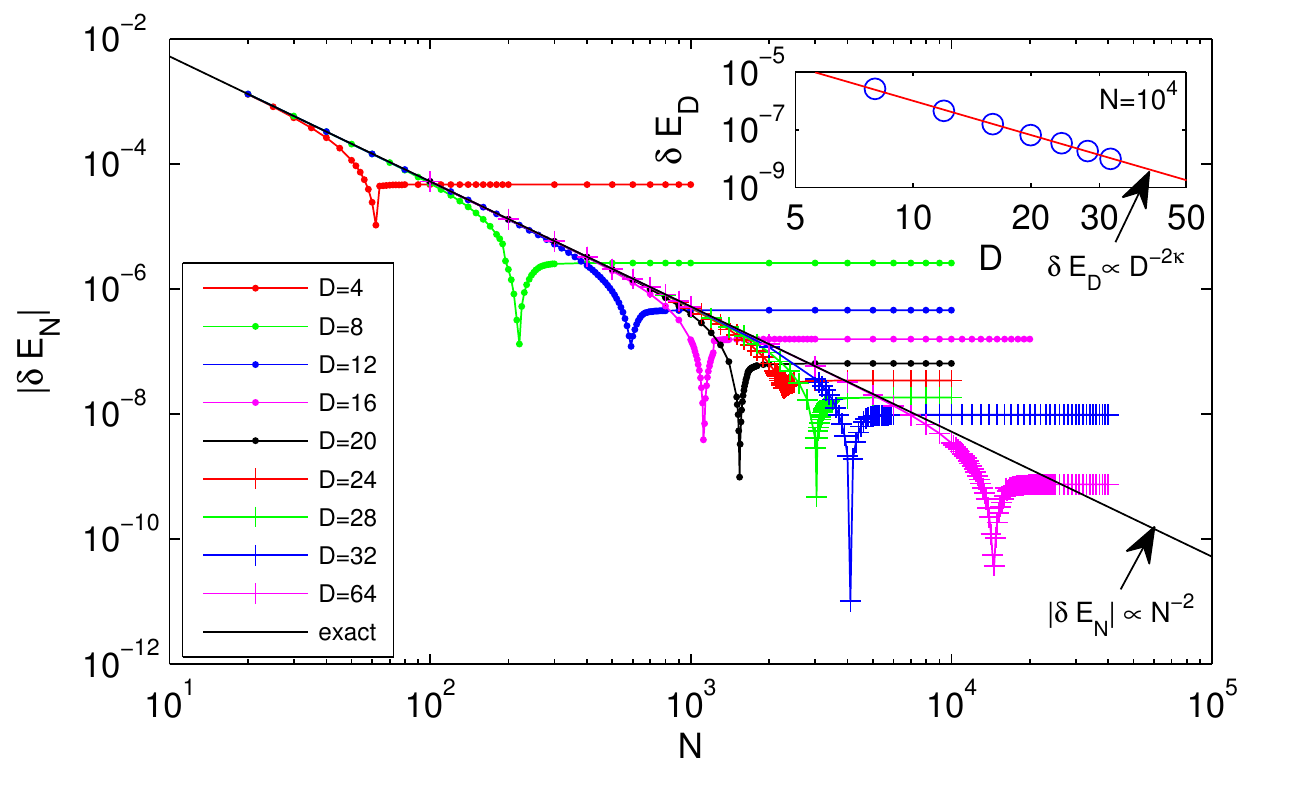}
  b)\includegraphics[width=1.0\columnwidth]{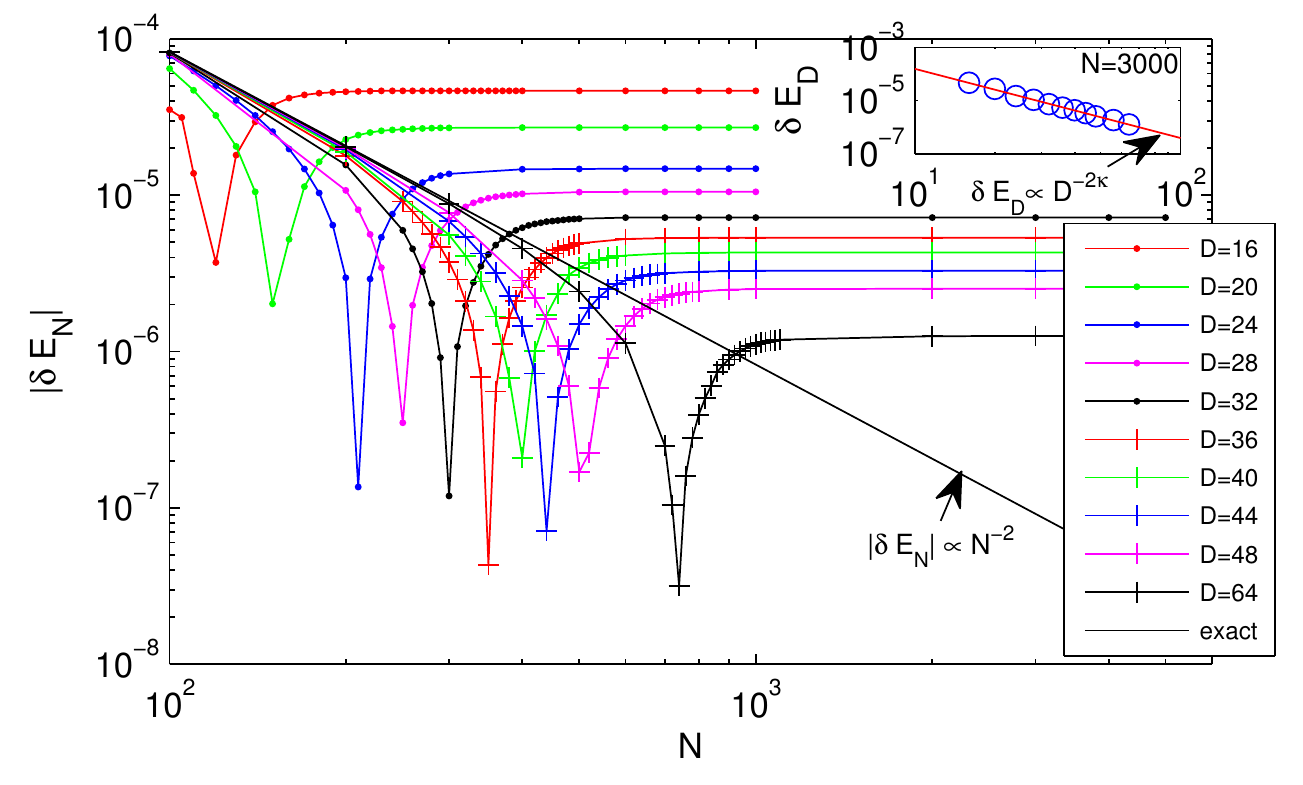}
  \caption{
    (Color online).
    a) Quantum Ising model, two regimes for simulations of PBC chains with
    lengths in the range $20 \le N \le 4\cdot 10^4$, and $D$ in the range
    $4 \le D \le 64$. Each line represents simulations performed at fixed
    $D$ and different $N$. The plots show the absolute value of the difference $\delta E_N = E_0(N,D=const.)-E_0^{\infty}$.
    The FSS is represented by a diagonal black line following the scaling
    from Eq. \ref{eq:CFT_Escaling_PBC}.
    All data sets initially follow this line.
    The FES regime corresponds to the various horizontal lines, where
    $\delta E_N$ saturates for different $D$ to different values
    $\delta E_D$ that do not depend on $N$. In the inset we collect these values to show that they reproduce the expected behavior of the FES.
    The two regimes are separated by the appearance of a pronounced peak.
    Since we plot an absolute value, the peak is nothing more than the
    change of sign in the difference $E_0(N,D)-E_0^\infty$ when moving
    from the FSS regime (\ref{eq:CFT_Escaling_PBC}) to the
    FES regime (\ref{eq:MPS_Escaling_TL}).
    b) The same plot for the HB model tells us that here the FSS is much
    more difficult to study, since all data-sets deviate very soon from
    the pure FSS prediction.}
\label{fig:E0_ND-E0_inf}
\end{figure}

The two regimes are very clearly distinguished in
Fig. \ref{fig:E0_ND-E0_inf} where we present plots of the absolute value
of the difference of the ground state energy obtained with MPS simulations
and the exact value in the thermodynamic limit

\begin{equation}\label{eq:delta_E_ND}
  \delta E_{N,D}=E_0(N, D)-E_0(\infty)
\end{equation}
\noindent
as a function of $N$ in a log-log scale.
Note that in Fig. \ref{fig:E0_ND-E0_inf} we make an abuse of notation
by using $\delta E_N = (\delta E_{N,D})_{D=const.}$ and
$\delta E_D = (\delta E_{N,D})_{N=const.}$, which can be only done as long
as we specify what the constant value of $D$ or $N$ is.
The data is collected from several simulations of the critical IS
with PBC for chain lengths in the range $20 \le N \le 4\cdot 10^4$
(panel a) and of the HB with PBC in the range
$10^2 \le N \le 5\cdot 10^3$ (panel b).
$D$ is going in both cases up to $D=64$. Each line in the main plot
represents simulations performed for different $N$ at fixed $D$.
The FSS predictions of Eq. \ref{eq:CFT_Escaling_PBC} are straight
lines plotted in black. For small $N$, each set of data follows the
prediction of Eq. \ref{eq:CFT_Escaling_PBC}, which is a clear signal
of the FSS regime. The maximal $N$ for which the FSS prediction
holds increases with growing $D$ as expected.
However, each set deviates at some big enough $N$ from the FSS prediction
to eventually stabilize to a value of the energy difference that only
depends on $D$. This is a clear footprint of the FES regime scaling, as
described in Eq. \ref{eq:MPS_Escaling_TL}. In order to confirm this we
have added to both panels insets where we have plotted
several values of $\delta E_D$ for large fixed $N$ as a function of
$D$ in a log-log scale. Similar plots in the thermodynamic limit
can be found in~\cite{me-2010-MPOR}.
The linear fits (red lines) in the insets yield
$\kappa_{IS}\approx 1.9776$ for $N=10^4$ respectively
$\kappa_{HB}\approx 1.3025$ for $N=3000$. These values are compatible
with the analytical values obtained for $N\to\infty$ in
\cite{moore-2008}, namely $\kappa_{IS}^{anal}\approx 2.03425$ and
$\kappa_{HB}^{anal}\approx 1.34405$ and thereby confirm the scaling of
Eq. \ref{eq:MPS_Escaling_TL}.

Note that we are able to obtain much better precision for the IS than
for the HB model at at the same computational cost .
This is visible by comparing the panel a) to the panel b) and observing
that for fixed $D$, the curves for the HB model deviate from the FSS
at much lower values of $N$ than the corresponding ones for the IS model.

\subsection{The transition between the two regimes}
\label{sec:transition}

In Figure \ref{fig:E0_ND-E0_inf} we can observe that for each line with
constant $D$, the FSS region is separated from the FES region by a well
distinguishable peak in the absolute value of $\delta E_N$.
We would now like to show that
this transition does not depend on the choice of the observable but that
it indicates a global change in the wave function.

To this end we can investigate the trace distance between the exact
ground state of a chain with $N$ sites and the MPS obtained from a series
of simulations with different $D$. We have chosen the step size in $D$
as small as possible, i.e. $\Delta D=1$.
Since the exact ground state wave function is only available for very small
systems due to the exponential scaling of the number of parameters,
we use as a \emph{reference state} a MPS approximation of the ground
state with very big $D$. For the $N$ range in question, the biggest
available bond dimension is $D=64$. Note that the energy difference between
the exact ground states and the \emph{reference states} is much smaller
than the difference to the MPS we want to compare to (see Fig.
\ref{fig:ISHB_Nall_Dall}).

\begin{figure}[ht]
  \begin{center}
   a)\includegraphics[width=1.0\columnwidth]{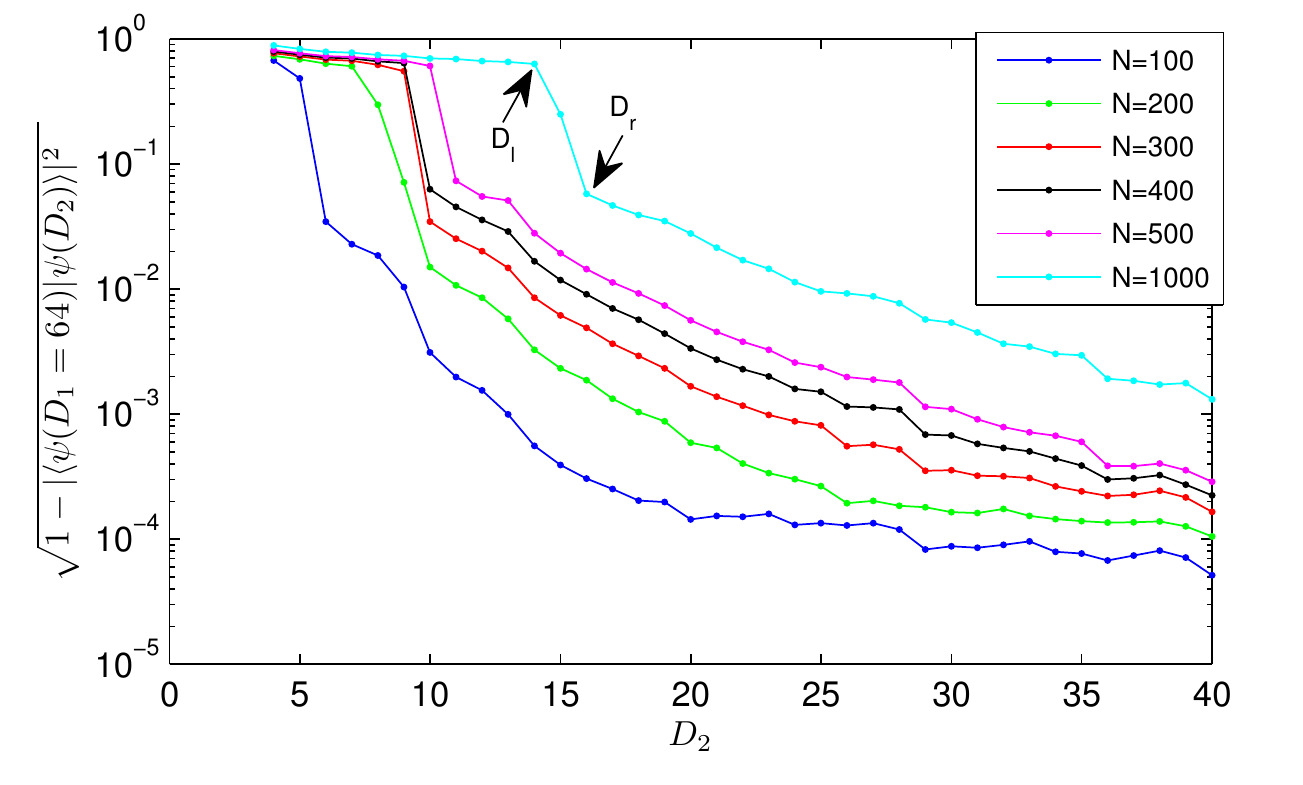}
   b)\includegraphics[width=1.0\columnwidth]{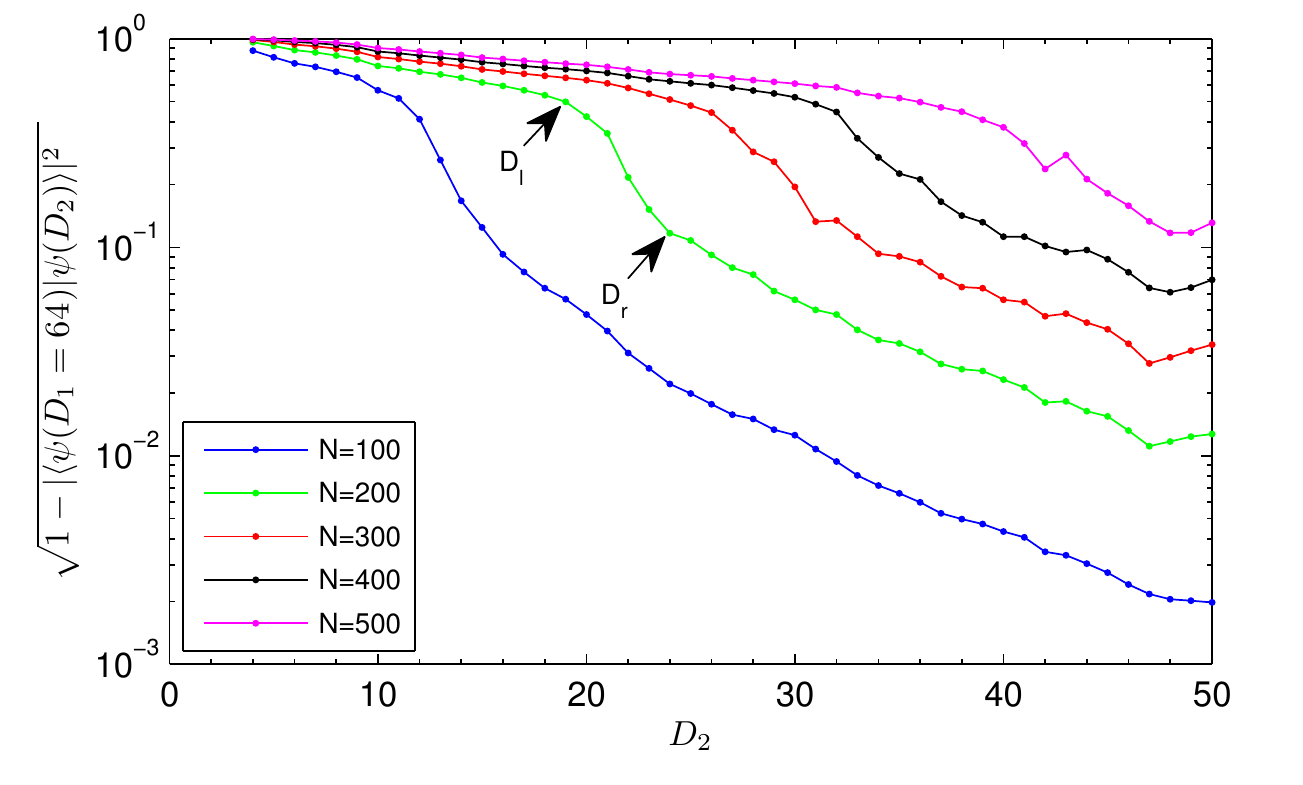}
\end{center}
  \caption{
    (Color online).
    a) Quantum Ising model:
    trace distance between \emph{reference states} with $D_1=64$
    and ground state MPS with bond dimensions $D_2$ for several
    different chain lengths.
    b) Heisenberg model:
    trace distance between \emph{reference states} with $D_1=64$
    and ground state MPS with bond dimensions $D_2$ for several
    different chain lengths.
  }
\label{fig:IS_TD_D64}
\end{figure}

Figure~\ref{fig:IS_TD_D64} shows the trace distance between states with
relatively small $D$ and \emph{reference states} for several different
chain lengths $N$ for both the IS and the HB models.
Note that for every $N$ there is a jump in the
trace distance between states that are very far away from the
\emph{reference state} and states that are at least one order of
magnitude closer to it. For the IS model the jump is very steep and
for each line of constant $N$ we can clearly identify $D_l$ as the
biggest $D$ in the left (FES)
regime and $D_r$ as the smallest $D$ in the right (FSS) regime.
In this case the appearance of the jump evidently indicates the transition
from the FSS regime, where the trace distance is close to zero to the
FES regime where the trace distance abruptly increases.
For the HB model the transition
is much smoother and we can not unambiguously define $D_l$ and $D_r$ for
all lines with constant $N$. This happens presumably due to the fact
that taking MPS with $D=64$ as \emph{reference states} is not accurate
enough in the case of the $HB$ model.

However, at least for the IS model, we are now in
the position to check if our intuitive expectation, that the transition
occurs precisely when the correlation length of the finite size MPS
reaches the size of chain as described in Eq. \ref{eq:transition},
is quantitatively correct. The correlation length of MPS with finite
$D$ reads according to Eq. \ref{eq:feg} as $\xi(D)=k_c \cdot D^\kappa$
where $k_c$ is a proportionality constant.
For our numerical study we obtain the parameters $k_c$ and $\kappa$ in
the appendix \ref{sec:eff_corr_len}. We can confirm that for each line
of constant $N$ the jump in the trace distance is consistent with our
assumption, i.e. $\xi(D_l) < N < \xi(D_r)$.

Furthermore we would like to mention that jumps also occur in other
quantities at the same $D_l$, for instance in the half-chain
correlation function reported in~\cite{me-2010-PBCI} (see figure $9$ in
that work for a plot of the jump for $N=500$).
The fact that the induced correlation length $\xi(D)$ in the FES is smaller
than the size of the system, suggests that the state is completely unaware
of the presence of the boundaries.
This confirms our intuition that MPS in the FSS regime are faithful
approximations of finite chains with PBC while MPS in the FES regime
do not capture properties related to the boundary conditions.

Summarizing, the main point of this section is that if one is interested
in the effects of PBC, results collected for $D$ smaller than $D_r$
should not be taken into account.
Note that due to the residual dependence on $D$
(see the appendix \ref{sec:correction_to_scaling} for details on this
point) one still has to extrapolate the results in the limit
$D\to\infty$ in order to obtain accurate results.
If, on the other hand, one is interested only in local universal
quantities (i.e. where boundary conditions are irrelevant), there is
no point in simulating the system with PBC and one should rather perform
a standard FES study \cite{tagliacozzo-2008}.

\subsubsection{The real scenario, a complex cross-over induced by corrections to the scaling}
\label{sec:correction_to_scaling}

We have seen above how in the extremal regions of Figure
\ref{fig:E0_ND-E0_inf}, the simulation results follow the behavior
predicted by FSS respectively FES. In the intermediate region however,
the simulations display a behavior that can not be attributed to
either regime. We would now like to point out that the real picture is
somewhat more complex than the two-regime interpretation given above.

The leading scaling behavior given in Equations \ref{eq:CFT_Escaling_PBC}
and \ref{eq:MPS_Escaling_TL} represents only the first terms of the series
expansion of more complex analytic corrections.
Thus these terms are accompanied by higher order terms called corrections
to scaling. In order to understand the scenario we must consider the
general Taylor expansion of a two variable function. Let us consider two
variables $\Delta _D$ and $\Delta _{N}$ with the property that
$\lim_{D\to \infty } \Delta _D =0$ and $\lim_{N\to \infty } \Delta_N =0$.
Obviously these variables can be identified with the gaps proportional to
the inverse of the correlation length induced by the system size $N$ and by
the finite matrix dimension $D$ as defined in Eq. \ref{eq:fsg} and Eq.
{eq:feg}. Part of the scaling ansatz consists in assuming that all
universal quantities are universal functions of these two variables.

Let us review the case of a one-scale problem. In this case, by
neglecting higher than quadratic terms in the vanishing variable
(e.g. $\Delta_n$) we get the following series expansion for some
universal function $g$

\begin{equation}\label{eq:one_scale}
  g_{\Delta_N} = g_{0} + \partial_{\Delta_N} g_{0} \Delta_N
  + \frac{1}{2} \partial^2_{\Delta_N} g_{0} \Delta_N^2  + \cdots
\end{equation}
\noindent
In the regime where $\Delta_N^2 \ll 1$, the first two terms are
considered the leading scaling behavior while the rest provides only
higher order corrections. If we now take a two scale problem

\begin{multline}
\label{eq:two_scales}
f_{\Delta_D, \Delta_N} = f_{0,0} + \partial_{\Delta_D} f_{00} \Delta_D + \partial_{\Delta_N} f_{00} \Delta_N \\
+ \frac{1}{2} \left(  \partial^2_{\Delta_D} f_{00}  \Delta_D^2+\partial^2_{\Delta_N} f_{00} \Delta_N^2\right) \\
+ \partial^2_{\Delta_D \Delta_N} f_{00} \Delta_D \Delta_N +\cdots
\end{multline}
\noindent
in the regime where $\Delta_D \ll \Delta_N^2 \ll \Delta_N$ we are back
to the previous situation and we can apply the one-scale ansatz of Eq.
\ref{eq:one_scale} to the function $g(\Delta_N)=f(\Delta_N, 0)$; the same
thing is valid in the opposite regime $\Delta_N\ll\Delta_D^2\ll\Delta_D$
with the obvious substitution $g(\Delta_D)=f(0, \Delta_D)$. These two
limits would correspond to what we have called in the main text
the FSS regime and the FES regime (see Eq. \ref{eq:fsr} and \ref{eq:fer}).

\begin{figure}[ht]
  \begin{center}
    \includegraphics[width=1.0\columnwidth]{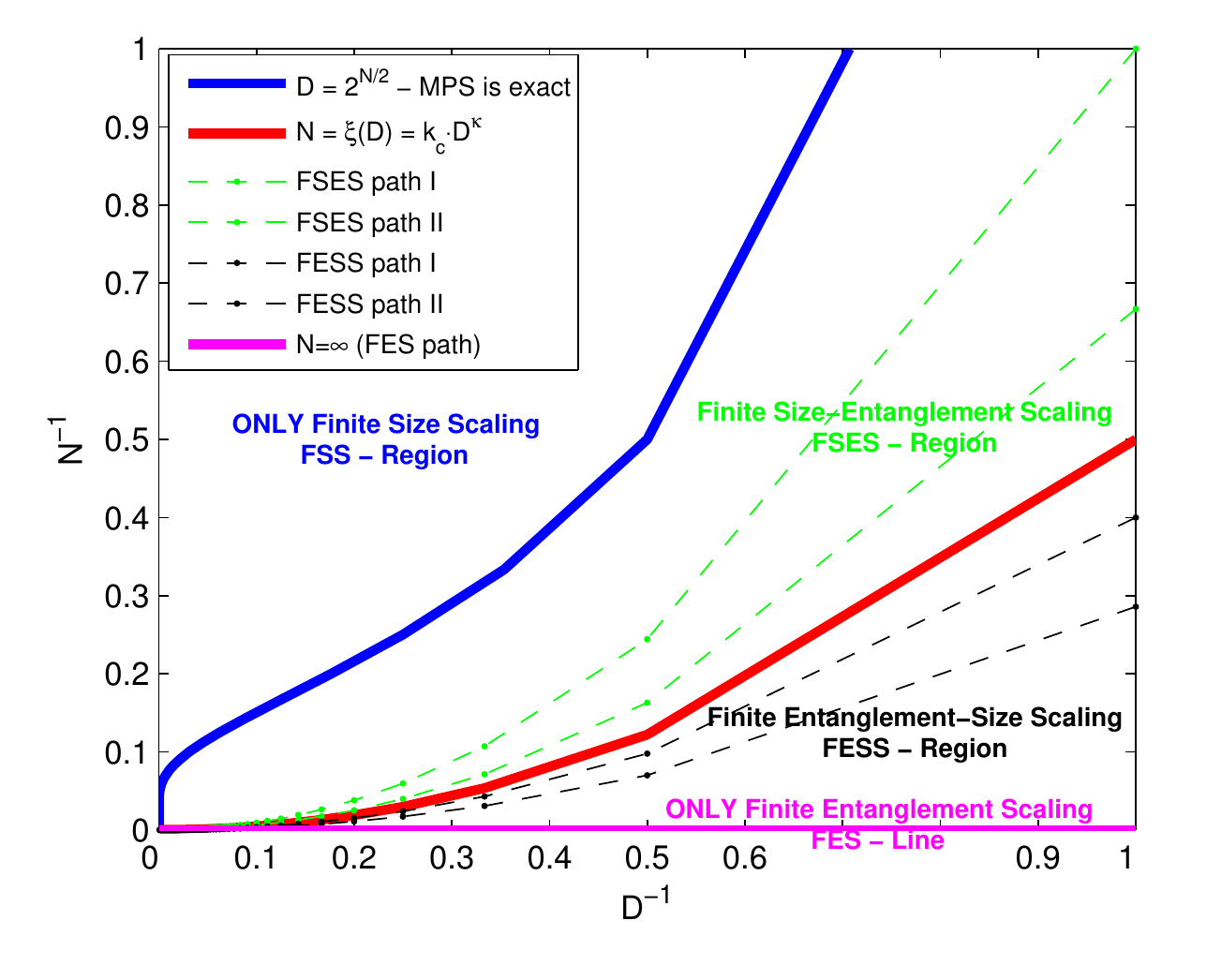}
  \end{center}
  \caption{
    (Color online).
    Classification of MPS simulations of spin chains according to
    the simulation parameter pair $\{N,D\}$.
    All lines ending in the origin denote possible paths to approach the
    thermodynamic limit of critical systems when doing MPS simulations.
  }
\label{fig:FSES_phase_diagram}
\end{figure}

Now in general, there is a huge regime where for example
$\Delta_N^2 \ll \Delta_D  \le \Delta_N$ or
$\Delta_D^2 \ll \Delta_N  \le \Delta_D$.
In this case the leading scaling behavior is modified by corrections
that are not proportional to the next power of the relevant variable but
to the ratio among the two variables. Indeed if we consider the
scenario where $\Delta_D^2 \ll \Delta_N  \le \Delta_D$ in Eq.
\ref{eq:two_scales} we obtain

\begin{equation}\label{eq:two_scales_cross}
  f_{\Delta_D, \Delta_N} = f_{0,0}
  + \Delta_D \left(\partial_{\Delta_D} f_{00}
  + \partial_{\Delta_N} f_{00} \frac{\Delta_N}{\Delta_D}\right)
  + \cdots
\end{equation}
\noindent
How relevant the correction is clearly depends on the scale separation,
i.e. on how close
$\Delta_N/\Delta_D$
is to one.
In the following we give a sketch of how this cross-over region looks
like and we introduce two new terms: \emph{Finite Entanglement-Size
Scaling} (FESS for) the region where the leading scaling is due to the
finite size of the matrices and the corrections come from the size of the
system and \emph{Finite Size-Entanglement Scaling} (FSES) where the
leading scaling is due to the size of the system and the corrections
come from the size of the matrices.

Figure~\ref{fig:FSES_phase_diagram} shows a classification of MPS
simulations according to the simulation parameter pair $\{N,D\}$.
The thermodynamical limit can be approached by moving along any
path towards the origin of the diagram $\{N^{-1}=0,D^{-1}=0\}$.
However in order not to distort the scaling analysis by mixing the
different $N$ and $D$ related corrections, moving from one
point to the next on the path should leave the ratio
$\Delta_N/\Delta_D$ unchanged. This is equivalent to the requirement
that any path is completely determined by the path constant
$k=N/D^\kappa$.

We can distinguish three different regions and three important lines
in Fig.~\ref{fig:FSES_phase_diagram}
In the region above the \emph{blue line} which is defined by
$D=d^{N/2}$, the MPS bond dimension is
large enough to represent the ground state exactly. Of course doing MPS
simulations in this regime is pointless since the computational cost
becomes exponential in $N$ and there is no advantage over exact
diagonalization. Thus no matter which path towards $N\to\infty$ we
choose in this region, it is completely equivalent to FSS.
The \emph{magenta line} with $N^{-1}=0$ represents the only path
along which pure finite entanglement scaling (FES) holds.
The \emph{red line} represents the path along which the induced
correlation length is equal to the system size, i.e. $N=\xi(D)$.
We will call this line in the following the \emph{critical line}.
For critical models without conformal invariance the
\emph{critical line} can be obtained
using the method described in appendix~\ref{sec:eff_corr_len}.
Between this line and the \emph{FES line} there is a region
where $N>\xi(D)$. All simulations done in this region barely registrate the
boundaries of the system and the fixed point MPS is more or less
the same like that of a $N=\infty$ simulation with same $D$.
However there is a slight effect due to the finite size for points close
to the $N=\xi(D)$ line as can be seen in Figure~\ref{fig:ISHB_Nall_Dall}.
This is why we call this region the finite entanglement-size scaling
(FESS) region: the entanglement scaling predominates, but there is a small
trace of finite size scaling behavior.
The region between the \emph{critical line} and the FSS-regime describes
MPS simulations where $\xi(D)<N$, which turn out to
reproduce faithfully the long range correlations throughout
the entire chain (see figure 9 in our previous work~\cite{me-2010-PBCI}).
The FSS aspect predominates in this region, however there is also
the inherent error of MPS simulations with $D<d^{N/2}$,
so we call it the finite size-entanglement scaling (FSES) region.

Despite the rigorous classification of regimes from Figure
\ref{fig:FSES_phase_diagram} we will restrict ourselves in the following
to discriminate merely between the regimes on different sides of
the critical line. We do this in order to improve the readability.
Thus we will refer to both FSS and FSES as FSS; analogously we will denote
both FES and FESS as FES.

\begin{figure}[ht]
  \begin{center}
  a)\includegraphics[width=1.0\columnwidth]{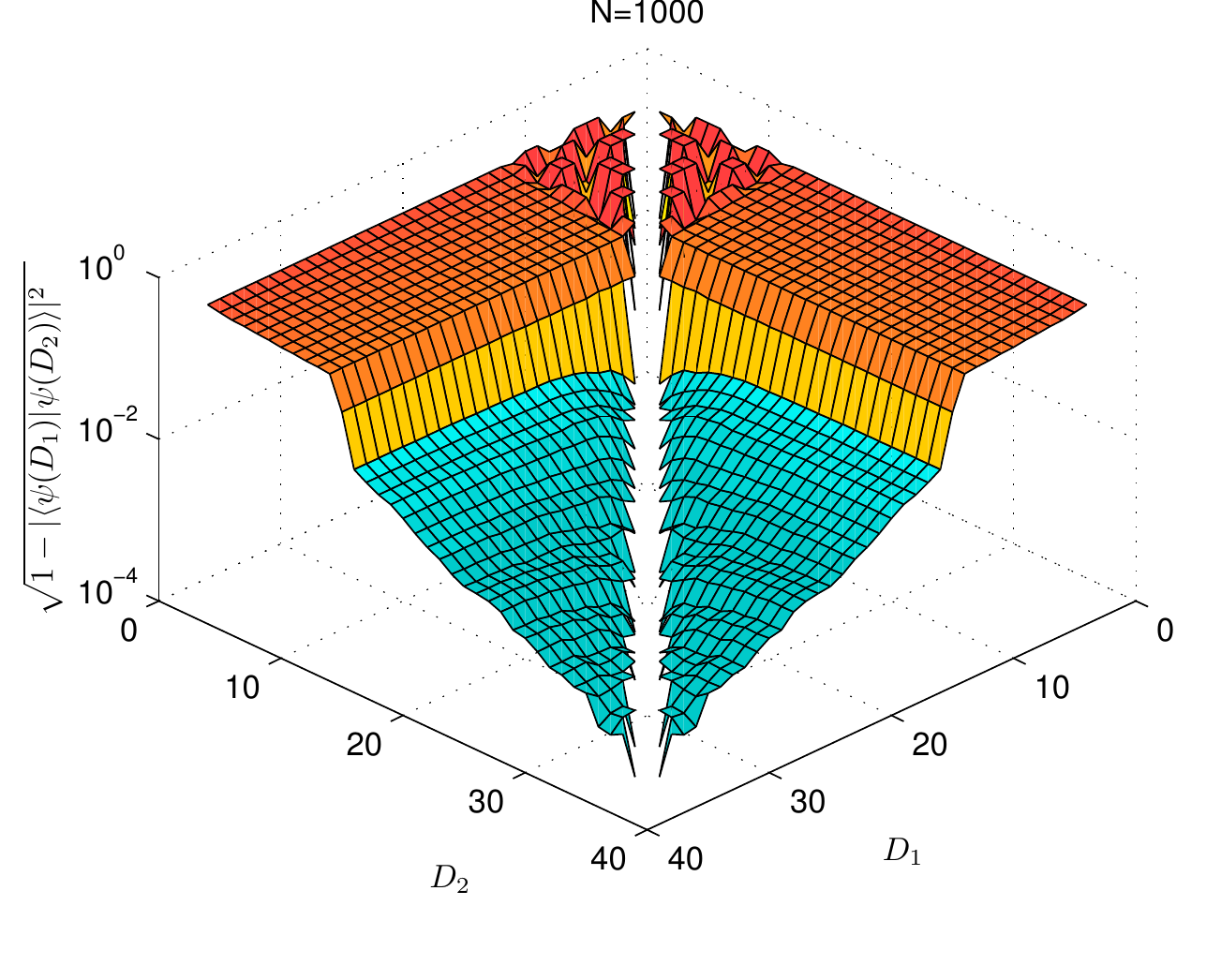}
  b)\includegraphics[width=1.0\columnwidth]{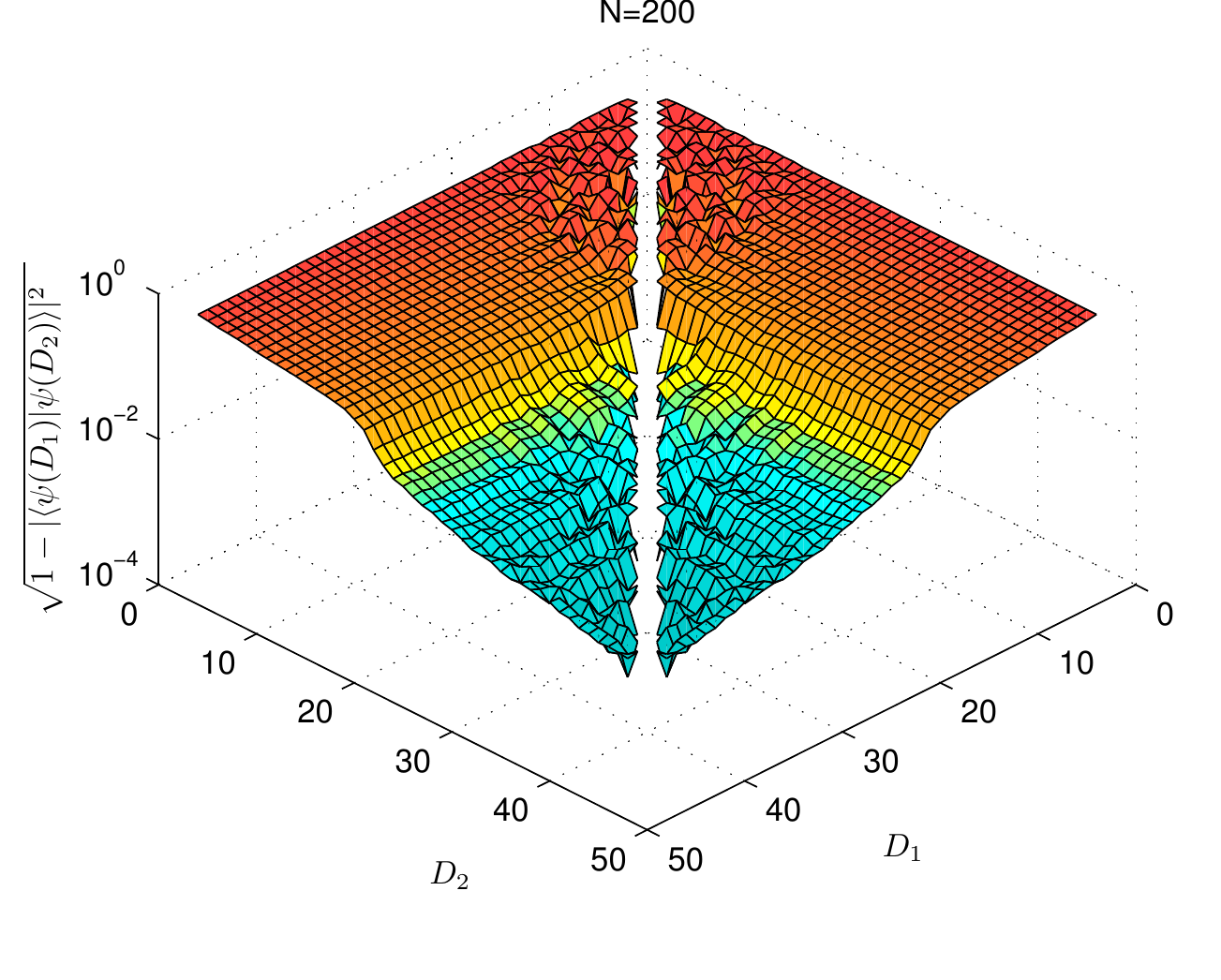}
   \end{center}
  \caption{
   a) (Color online).
    Quantum Ising model:
    trace distance between ground state MPS with different
    bond dimensions $D_1$ and $D_2$ for a chain with $N=1000$ sites.
  b)  Heisenberg model:
    trace distance between ground state MPS with different
    bond dimensions $D_1$ and $D_2$ for a chain with $N=200$ sites.
By using relatively small bond dimensions we are able to localize the transition between the FES and FSS regime  for each $N$. This can be used in case  we are interested in performing a FSS analisys by providing  the lower bound $D_r$ such that the state we obtain is not orthogonal to the exact state.
 }
\label{fig:IS_TD_D1D2_N1000}
\end{figure}

\subsection{Minimal $D$ for faithful simulations}
\label{sec:minimal_D_faithful_simulations}

We can  outline a direct application of the presence of a transition
between a FSS regime and a FES regime.
Suppose that we want to simulate a critical chain with PBC such that
it is in the FSS regime, i.e. the properties due to the boundary
conditions are faithfully reproduced since $N<\xi(D)$.
In order to minimize the computational cost we would like to use
the smallest possible $D$ that captures the PBC topology.
By looking at figure~\ref{fig:IS_TD_D64} it is clear that we would have
to choose $D=D_r$ to this end. The problem is that in order to make that
plot we had to use as \emph{reference states} MPS with very large $D=64$,
which is exactly what we would like to avoid in this case.
Fortunately it turns out that even without a large $D$ simulation it is
possible to detect the optimal $D=D_r$. This is due to the fact that all
MPS with $D\geq D_r$ have a much smaller trace distance among eachother
than with MPS with $D<D_r$.

The trace distance among all states with $D<40$ for a IS
chain with $N=1000$ is shown in figure~\ref{fig:IS_TD_D1D2_N1000} a).
The plot is of course symmetric in $D_1$ and $D_2$ and we have
omitted the points on the diagonal since they are trivially $0$.
The transition between the FSS and the FES regime is clearly
distinguishable at the same location of the jump as in figure~\ref{fig:IS_TD_D64}
but in this plot we used only MPS with relatively small bond dimension.
Note furthermore that if $D_1,D_2<D_r$ the trace distance
between these states is wildly oscillating. However if $D_1$ and
$D_2$ are on different sides of the jump, profiles similar
to figure~\ref{fig:IS_TD_D64} emerge.
Now it is clear how we can find the optimal $D=D_r$ with the smallest
computational cost possible: for a given $N$ run the PBC simulations
by increasing $D$ in small steps, ideally $\Delta D=1$. After each
simulation compute the overlap with all previously obtained MPS and when
the nice profile with the jump appears, we know we have reached $D=D_r$.
The same strategy can be employed for the HB model, however, just like in
figure \ref{fig:IS_TD_D64}, the transition is much smoother in this case.

As a side remark note that due to the fast decay of the eigenvalues of
the MPS transfer matrix one can compute the overlaps with computational
cost scaling like $O(nD^3)$. The meaning of $n$ and the method how to
achieve this is described in~\cite{me-2010-PBCI}.

\subsection{Thermodynamic limit of the transition}
\label{sec:IS_TL}

What can figure~\ref{fig:IS_TD_D64} tell us about the behavior of
the transition between the FSS and FES in the limit $N\to\infty$?

For the IS model, qualitatively the height of the jump seems to
remain constant for increasing $N$.
The trace distance between MPS with bond dimensions $D_l$ and $D_r$
to the \emph{reference state} also seems to remain more or less stable
but this is of course not enough evidence for the persistence
of the transition in the thermodynamic limit.
In appendix \ref{sec:TL_of_transition}, we present a detailed analysis
that shows that for the
IS model i) the $N\to \infty$ limit of the trace distance between the
exact ground state (approximated by a \emph{reference state})
and MPS obtained in the FSS regime is strictly
bigger than zero, ii) the same limit for the trace distance with
respect to MPS obtained in the FES regime is zero. ii) implies that
states in the FES regime are globally orthogonal to the exact ground state
of the PBC chain. As we already mentioned above this does not affect the
possibility to extract local universal information from those states.
However ii) clearly shows that MPS in the FES regime are globally not a
good approximation for the ground state of the IS model with PBC.

Unfortunately we cannot obtain the same conclusions for the HB
model. Presumably this is due to the fact that the \emph{reference states}
that we use are not a good enough approximation of the true ground state
of the model in this case. This becomes clear if we look again at
figure~\ref{fig:ISHB_Nall_Dall}:
for the IS model the $D=64$ states have a much better precision than the
MPS we compared them to in order to prove the persistence of the transition
in the thermodynamic limit (see Appendix \ref{sec:TL_of_transition} for
details). For the HB model on the other hand, the $D=64$ line covers
almost three orders of magnitude in the relative precision plot;
at its maximum it is over one order of magnitude above the points
belonging to MPS that we must compare the \emph{reference states} to
in order to perform our
analysis of the thermodynamic limit (e.g. the data points with $N=100$
and $D=48$). The $D=128$ line in right plot
of figure~\ref{fig:ISHB_Nall_Dall} seems to fulfill similar
requirements like the $D=64$ line in the left plot.
However, in that regime, for $N\ll\xi(D)$, the PBC algorithm is very
inefficient and it would take unreasonably long to obtain the data points
for $D=128$.

\subsection{The scaling function}
\label{sec:scaling_function}

\begin{figure}[ht]
  \begin{center}
   a)\includegraphics[width=1.0\columnwidth]{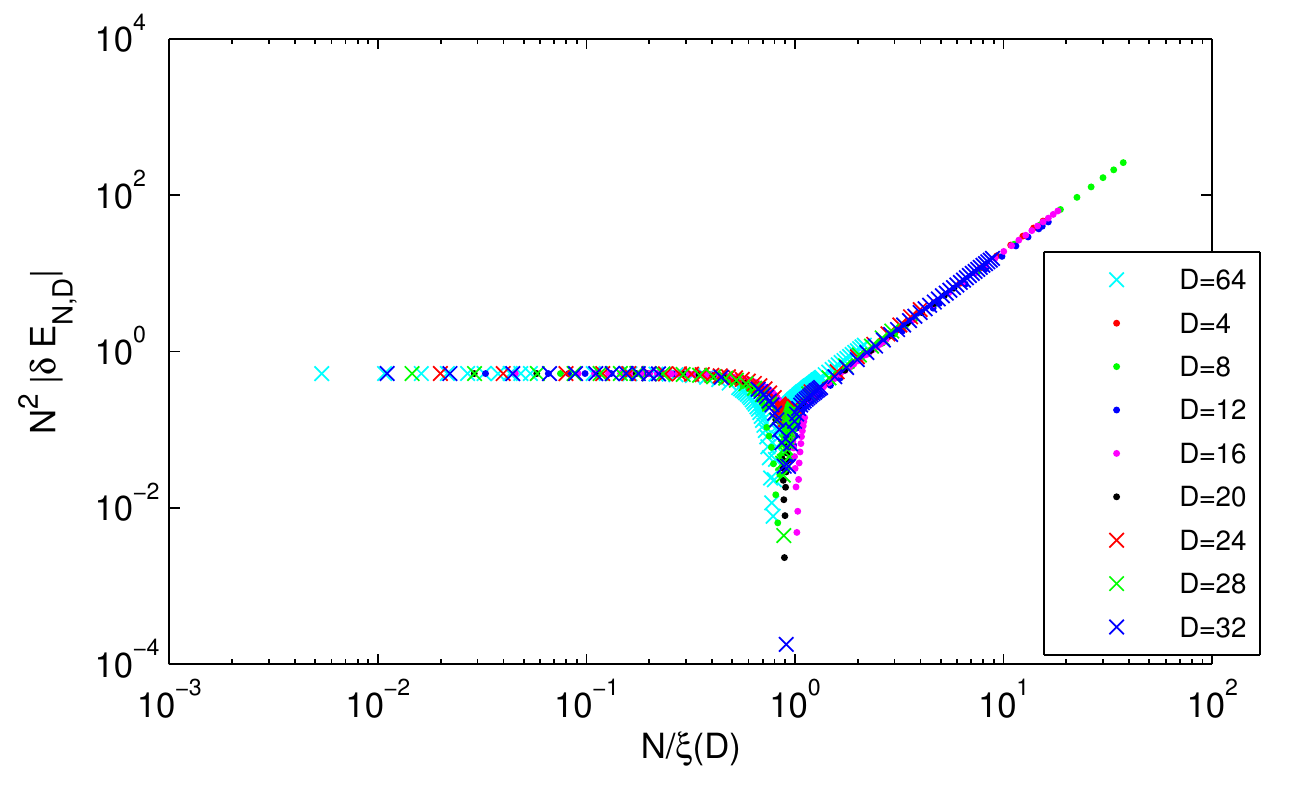}
   b)\includegraphics[width=1.0\columnwidth]{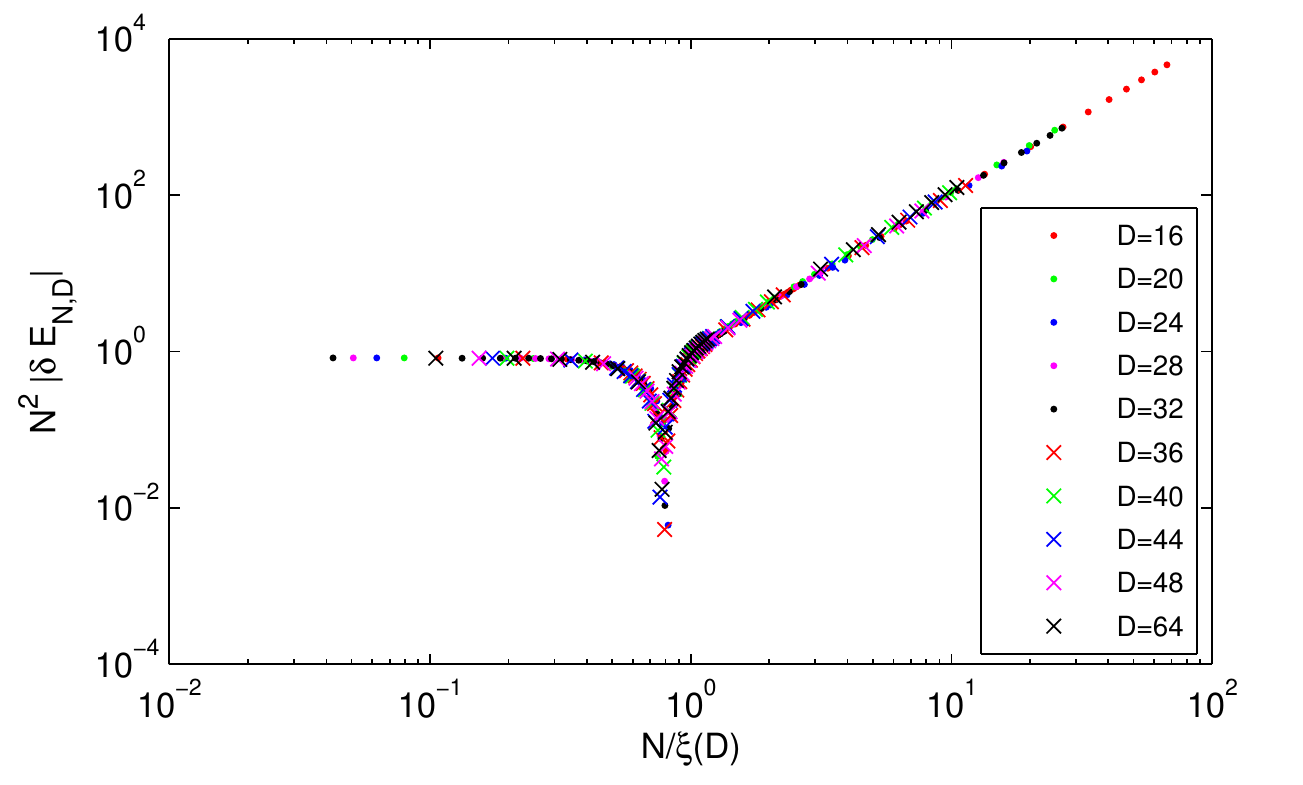}
\end{center}
  \caption{
    (Color online).
    a) Quantum Ising model:
    log-log plot of $N^2 \lvert \delta E_{N,D} \rvert$ versus
    $N / \xi(D)$ that illustrates the collapse of the data into a single
    curve. The points with $D=64$ slightly deviate from the curve
    traced by data points with smaller $D$.
    b) Heisenberg model:
    log-log plot of $N^2 \lvert \delta E_{N,D} \rvert$ versus
    $N/\xi(D)$ that illustrates the collapse of the data into a single
    curve.   }
\label{fig:IS_collapse}

\end{figure}

Finally we conduct an analysis of the scaling of MPS simulations
across the entire interval $N/\xi(D)\in (0,\infty)$ which covers all
possible pairs $\{N,D\}$.
This is very much in the spirit of the scaling analysis performed
by Nishino et al. for classical 2D systems in Ref.~\cite{nishino-1996}.
The main differences are that in our case the energy difference
$\delta E_{N,D}$ can take both positive and negative values, and
that we obtain the effective correlation length $\xi(D)$ from an analysis
of the \emph{humps} in the relative precision of the energy
(see Appendix \ref{sec:eff_corr_len} for details) instead of using
the ratio between the two biggest eigenvalues of the MPS transfer matrix.

Analogously to Nishino, we first eliminate the FSS scaling from
$\lvert \delta E_{N,D} \rvert$ and then we plot the result
(in our case this is $N^{2} \lvert \delta E_{N,D} \rvert$) as a function
of $N/\xi(D)$. The fact that all data (with exception of the
$D=64$ points for the IS model) collapses into a single curve justifies
the assumption that

\begin{equation}\label{eq:MPS_Escaling_2param}
  \delta E_{N,D}=E_0(N,D)-E_0(\infty)=\frac{f(N/\xi(D))}{N^2}
  \,\,\,.
\end{equation}
\noindent
with some scaling function $f(x)$ that is not exactly known.
What we can easily write down however is its asymptotic behavior

\begin{equation}\label{eq:asymptotic_scaling}
\begin{split}
  \lim_{x\to 0}f(x)&=-\frac{v_f \pi c}{6}\\
  \lim_{x\to \infty}f(x)&=\Delta \cdot \Big( \frac{N}{D^\kappa} \Big)^2
  \,\,\,.
\end{split}
\end{equation}
\noindent

For the IS model we have used for $\xi(D)$ the expression obtained from
the hyperbola fit in Fig.~\ref{fig:ISHB_effCORRlen} of the Appendix
\ref{sec:eff_corr_len}, i.e. $\xi(D)=3.810\cdot D^{2.042}$.
Note that in plot a) of the Fig.~\ref{fig:IS_collapse}, the data
for different $D$ collapses almost perfectly in the extremal regimes
$N\ll\xi(D)$ and $N\gg\xi(D)$. There is a slight deviation of
the $D=64$ curve that can be explained if we look at figure 1
in~\cite{me-2010-MPOR} (there the $D=64$ data point also slightly
deviates from the line that is traced by the points with $D<64$).
In the regime where $N\approx\xi(D)$ the curves do not collapse so
nicely which is a manifestation of the fact that the \emph{humps}
in Fig.~\ref{fig:ISHB_Nall_Dall} are so different for
the IS model.

For the HB model we have used for the effective correlation length
$\xi(D)=3.647\cdot D^{1.338}$ as obtained from the hyperbola fit
in Fig.~\ref{fig:ISHB_effCORRlen}.
Plot b) of Fig.~\ref{fig:IS_collapse} shows an almost perfect collapse
even in the region $N\approx\xi(D)$. Presumably this is due to the
fact that for the HB model the \emph{humps} in Fig.
\ref{fig:ISHB_Nall_Dall} are much more similar among
each other than in the case of the IS model.


\section{Conclusions}
\label{sec:conclusions}

An accurate analysis of MPS simulations of critical spin chains with PBC
reveals the appearance of two regimes. The FSS regime where the energy gap of
the system is induced by the size of the system and the FES regime where
an effective energy gap  is induced by finite $D$.
While in both regimes local universal quantities can be extracted by
studying the scaling of the observable with respect to the
relevant variable (the size of the system for the FSS or the size of the
MPS matrices for the FES regime), we have shown that for the Quantum Ising
model, states in the FES regime are orthogonal to the exact ground state
in the thermodynamic limit. Intuitively this happens due to the
fact that for MPS simulations in the in the FES regime, the induced
correlation length is smaller than the system size and thus the MPS is
not aware of the size of the system.
Since in critical systems the boundary conditions strongly affect global
properties of the system, this result seems quite natural.

Our results can be interpreted as a further benchmark for recently
introduced algorithms that try to lower the computational cost of PBC
simulations with MPS
\cite{pippan-2010,me-2010-PBCI, zhou-2009, rossini-2011}
(see Appendix~\ref{sec:other_algorithms}).
Here we provide strong hints that in order to correctly describe the
ground state of a finite chain with PBC for critical systems, these
algorithms should be used with care in order not to obtain wave functions
that are orthogonal to the exact ones. What one would indeed interpret as
the MPS tensor for a PBC chain, in some regime could turn out to be closer
to the MPS tensor of an infinite OBC system.

However, considering that for OBC systems the approach to the
thermodynamic limit is by no means slower than for PBC systems (for both
the ground state energy converges to the thermodynamic limit as a function
of the appropriate correlation length like $\xi^{-2}$), our results can be
also used in a constructive way. In order to extract universal information
about local operators, one is better off by using FES rather than FSS,
since simulations in the FES regime have a much better scaling of the
computational cost.

Things are more complex if one is interested in global observables, such as e.g. two point correlation functions at half chain length. For PBC systems
the scaling analysis must be performed in this case on paths with constant
$k=N/D^{\kappa}$ that lie completely in the FSS regime.
The computationally least expensive such path is the one where for every
given $N$, the MPS bond dimension $D$ is just big enough such that
$\xi_D > \xi_N$. We have shown how that minimal $D$ can be found for any
$N$ by looking at the overlap between MPS with increasing $D$ until
the discrete transition between the FES and the FSS regime is detected.
Regarding the scaling exponent $\kappa$, we have been able to numerically
confirm the theoretically predicted values with an accuracy of
approximately $0.4\%$ for both the Quantum Ising and the Heisenberg models.
Furthermore we have shown in Appendix~\ref{sec:eff_corr_len_analytical}
how the analytical expression for $\kappa$, originally derived in
\cite{moore-2008}, can be obtained in an alternative way.

Following Nishino's analysis for 2D classical systems~\cite{nishino-1996}
we have shown that also for MPS simulations of 1D quantum systems
the scaling of the MPS ground state energy in simulations with finite
$N$ and $D$ obeys a two-parameter scaling function. Finding an analytical
expression for this function is something that still has to be done.

A further interesting future line of research is to understand how to
extract information about the operator content of the Conformal Field
Theories related to the infrared behavior of the studied critical spin
systems (that strongly depend on boundary conditions
\cite{bloete-cardy-nightingale-1986,evenbly-pfeifer-luca-guifre-2010})
directly out of the MPS tensors.

\section{Acknowledgements}

We thank V. Murg, I. McCulloch, E. Rico, V. Eisler for valuable
discussions. The TDVP computation was kindly performed by J. Haegeman.
This work was supported by the
FWF doctoral program Complex Quantum Systems (W1210)
the FWF SFB project FoQuS, the ERC grant QUERG, and the ARC grants
FF0668731 and DP0878830. We aknowledge the financial support of the Marie Curie project FP7-PEOPLE-2010-IIF ``ENGAGES'' 273524.




\begin{appendix}

\section{Effective correlation length}
\label{sec:eff_corr_len}

\subsection{Analytical results}
\label{sec:eff_corr_len_analytical}

Recently it has been shown numerically that any MPS simulation of an
infinite spin chain leads to the emergence of an effective correlation
length induced by the finite rank $D$ of the MPS matrices,
even if the studied system is critical~\cite{tagliacozzo-2008}.
In Ref.~\cite{moore-2008} the authors relate the numerical observation
that $\xi(D)\propto D^\kappa$ from Ref.~\cite{tagliacozzo-2008}
to analytical results on the spectrum of the MPS transfer matrix
\cite{calabrese-2008} and to well-known results from conformal
field theory~\cite{bloete-cardy-nightingale-1986,affleck-1986}
in order to derive an analytical expression for the exponent $\kappa$.
Here we derive the same results in a different way.

The starting point for our argument is the same like the one in
\cite{moore-2008}, namely that corrections to the exact ground state
energy in the thermodynamic limit can have different origins.
On one hand conformal invariance yields in the vicinity of the critical
point (i.e. $\epsilon=|\lambda-\lambda_{crit}|/\lambda_{crit}\ll 1$)
according to Refs.
\cite{bloete-cardy-nightingale-1986,affleck-1986,moore-2008}

\begin{equation}\label{eq:Escaling_CFT}
  E_0(\xi_\epsilon)=E_0(\infty)+\frac{A}{\xi_\epsilon^2}
\end{equation}
\noindent
where $A$ is a non-universal constant. On the other hand, MPS
simulations with finite $D$ yield according to Refs.
\cite{tagliacozzo-2008,calabrese-2008,moore-2008}

\begin{equation}\label{eq:Escaling_MPS}
  E_0(\xi_D)=E_0(\infty)+\frac{\beta}{\xi_D} P_r(b,D)
\end{equation}
\noindent
where $\beta$ is a non-universal constant, $P_r(b,D)$ is the residual
probability due to the usage of finite $D$ and $b$ is related to the
dominant eigenvalue of the reduced density matrix of the half-chain (see
Refs.~\cite{calabrese-2008,moore-2008}).
Now it has been observed that the usage of finite $D$ in MPS simulations
close to the critical point leads to an effective shift of the
critical point (see Fig. 2 in~\cite{tagliacozzo-2008}). This
observation led us to the idea of equating the corrections
in Eqs. (\ref{eq:Escaling_CFT}) and (\ref{eq:Escaling_MPS})
and identifying $\xi_\epsilon$ with $\xi_D$.
Together with the assumption $\xi_D=k_c\cdot D^{\kappa}$ this yields

\begin{equation}\label{eq:kappa_derivation_01}
  P_r(b,D) = \frac{A}{\beta\cdot\xi_D} = A'\cdot D^{-\kappa}
\end{equation}
\noindent
where we have collected all constants into
$A'=A/(k_c\cdot\beta)$.

In the large $D$ limit (required due to our assumption of working in
the scaling limit), the residual probability reads according to
\cite{moore-2008}

\begin{equation}\label{eq:Pr_01}
  P_r(b,D) = \frac{2 b e^{-b} D}{\log D - 2b}
             e^{-(\log D)^2/4b}
\end{equation}
\noindent
where

\begin{equation}\label{eq:b_01}
  b = \frac{c}{12} \log \xi_D
    \approx \frac{c\kappa}{12}\log D
\end{equation}
\noindent
and $c$ denotes the central charge in the associated conformal
field theory. Inserting (\ref{eq:Pr_01}) and (\ref{eq:b_01}) into
(\ref{eq:kappa_derivation_01}) yields after several steps

\begin{equation}\label{eq:kappa_derivation_02}
  \frac{c\kappa}{6-c\kappa}
  D^{-\frac{c\kappa}{12}-\frac{3}{c\kappa}+1}
  = A'\cdot D^{-\kappa}
  \,\,\,.
\end{equation}
\noindent
Equating the exponents in (\ref{eq:kappa_derivation_02}) yields a
quadratic equation for $\kappa$ with the solutions

\begin{equation}\label{eq:kappa_derivation_03}
  \kappa_{\pm} = \frac{6}{c\cdot(1\pm\sqrt{\frac{12}{c}})}
  \,\,\,.
\end{equation}
\noindent
The physical root is the one that is positive for all values of $c$, i.e.

\begin{equation}\label{eq:kappa_derivation_04}
  \kappa = \frac{6}{c\cdot(1+\sqrt{\frac{12}{c}})}
\end{equation}
\noindent
which is exactly the result obtained in Ref.~\cite{moore-2008}.

\subsection{Numerical results}
\label{sec:eff_corr_len_numerical}

\begin{figure*}[ht]
  \begin{center}
    \includegraphics[width=1.0\textwidth]{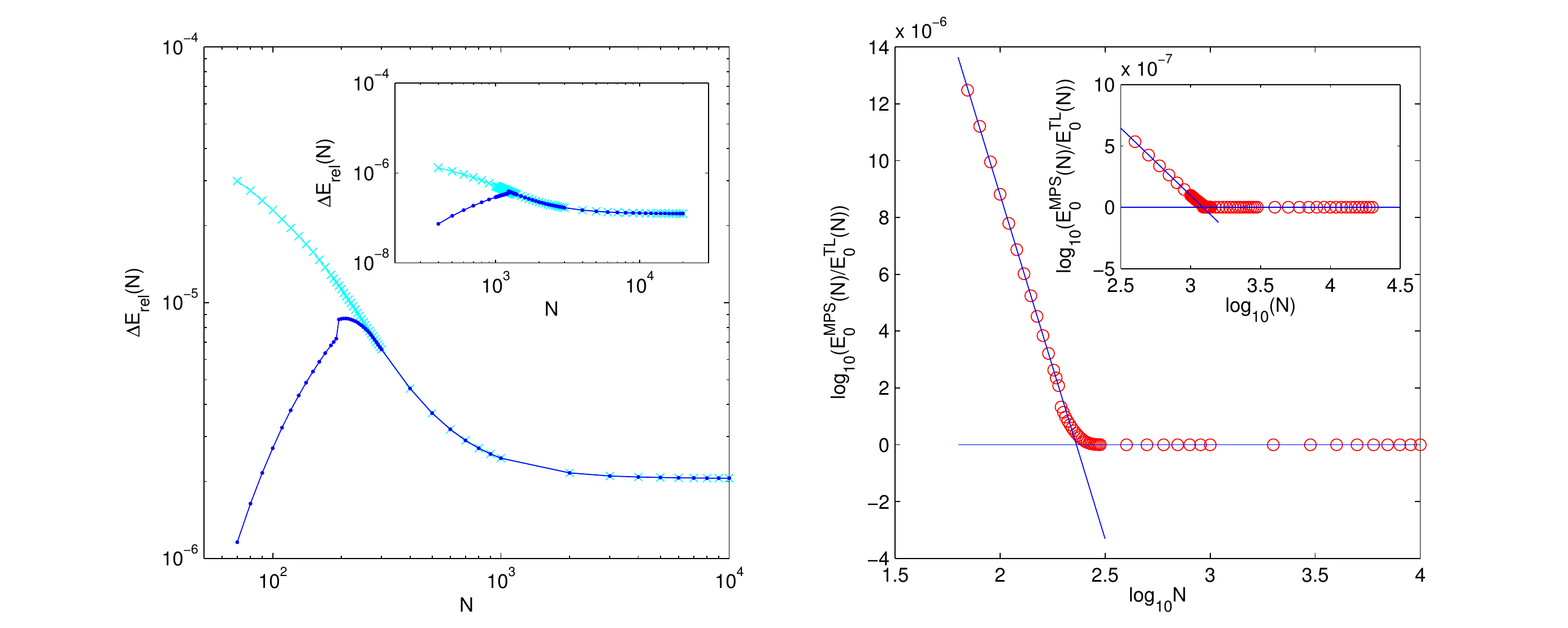}
  \end{center}
  \caption{
    (Color online).
    Quantum Ising model data.
    Left: relative precision of the PBC-MPS ground state energy (blue/dark)
    and relative precision of the state used as an input for the PBC
    algorithm (i.e. obtained by inserting the TL-MPS, cyan/bright)
    as compared to the exact result for $D=8$ and $D=16$ (inset).
    Right: ratio between the PBC-MPS energy and the one of the input
    state (TL-MPS) for $D=8$ and $D=16$ (inset). Fitting a degenerate
    hyperbola in form of two straight lines yields a well defined
    point (the intersection point) whose value as a function of $D$ is
    proportional to the effective correlation length.
  }
\label{fig:IS_Nall_D8D16_2linefit}
\end{figure*}

In this appendix we show how the effective correlation length $\xi(D)$
emerges in our simulations of finite spin chains with PBC.
As the scaling of the algorithm~\cite{me-2010-PBCI} is quasi-independent
of the chain length $N$ we can use it to approximate ground states of
arbitrary long chains with PBC.
The relative precision of the MPS ground state energy for
a given $D$ is plotted in figure~\ref{fig:ISHB_Nall_Dall}
as a function of $N$.
Each of the lines contains a \emph{hump} which can be interpreted as the
evidence for a finite correlation length $\xi(D)$.
In order to see this let us have a look at how the \emph{hump} emerges.
The left part of figure~\ref{fig:IS_Nall_D8D16_2linefit} shows
a comparison between the relative precision of the PBC-MPS ground state
energy (i.e. the MPS towards which the algorithm
in~\cite{me-2010-PBCI} has converged)
and the relative precision of the energy for the MPS that we
had used as a starting point for the gradient search.
As explained in~\cite{me-2010-PBCI} this is the local MPS tensor obtained
by imaginary time evolution~\cite{me-2010-MPOR} for a chain in the
thermodynamic limit (TL) when it is used in the finite PBC geometry.
One can see that for a given $D$, on the left side of the \emph{hump} there
is considerable improvement in the precision of the energy between
starting and ending point of the gradient search.
As one approaches the \emph{hump} from the left,
the improvement decreases in order to vanish completely on the right
side. This can be interpreted as follows: if $N$ is too large for
a given $D$, the finite chain looks for a local MPS-tensor
as if it would be infinite. Sites that lie further apart
than a certain correlation length $\xi(D)$ effectively do not see each
other. The transition to this region happens more or less smoothly
since for growing powers of the MPS transfer matrix $T$,
the subspace spanned by these powers gets smoothly restricted
to the dominant eigenvector
i.e. $T^N|_{N\gg\xi(D)}\approx\lambda_1^N\ket{\lambda_1}\bra{\lambda_1}$.

\begin{figure}[ht]
  \begin{center}
    \includegraphics[width=1.0\columnwidth]{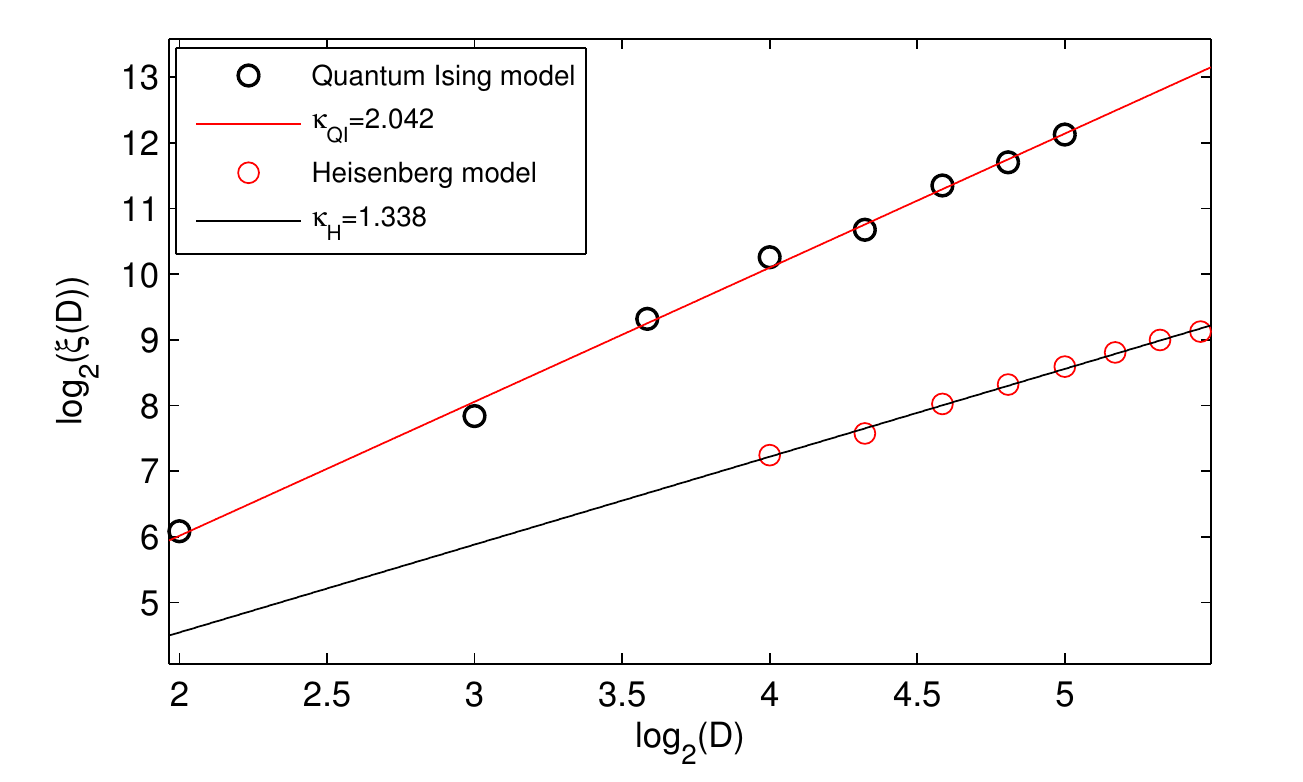}
  \end{center}
  \caption{
    (Color online).
    Quantum Ising (upper) and Heisenberg model (lower):
    linear fit of the logarithm of the effective correlation length
    as a function of the bond dimension $D$.
  }
\label{fig:ISHB_effCORRlen}
\end{figure}

Thus the \emph{humps} must represent some evidence for the emergence
of a finite correlation length, but how can we extract some reliable
numbers from them, as they differ considerably in shape and width?
The answer is given by the right part of
figure~\ref{fig:IS_Nall_D8D16_2linefit}. We have observed empirically
that if we make a log-log plot of $E_{0}^{final}(N)/E_{0}^{initial}(N)$
\footnote{
  $E_0^{final}(N)$ is the energy of the MPS obtained as the ground state
  by our conjugate gradient search for the finite chain with $N$ sites
  and PBC.
  $E_0^{initial}(N)$ is the energy obtained by plugging the MPS resulting
  from imaginary time evolution of the infinite chain into
  the finite size geometry with PBC.
  As mentioned in the main text, this MPS is used as the starting point
  of the conjugate gradient search.
}
we obtain approximately two straight lines connected by a small piece
that is more or less smooth. This picture is reminiscent of a rotated
hyperbola. We know furthermore that in the large $N$ limit all points
have ordinate $0$. This suggests to fit a hyperbola that is degenerated
to two straight lines through our data. The intersection of these lines
is a well defined point which should be proportional to $\xi(D)$.

Figure~\ref{fig:ISHB_effCORRlen} shows log-log plots of the effective
correlation length as defined above for both the Quantum Ising and
the Heisenberg models. After fitting straight lines through each of the
data sets we can read off the scaling $\xi(D)=k_c\cdot D^{\kappa}$
with $\{\kappa\approx 2.042, k_c\approx 3.810\}_{IS}$
and $\{\kappa\approx 1.338, k_c\approx 3.647\}_{HB}$.
Comparison with the analytical results
(i.e. $\kappa_{IS}^{anal}\approx 2.03425$ and
$\kappa_{HB}^{anal}\approx 1.34405$) yields a difference of roughly
$0.4\%$ for the Quantum Ising model and of roughly
$0.43\%$ for the Heisenberg model. These results are the ones
we refer to in Sec.~\ref{sec:transition} as the ones fulfilling
$\xi(D_l)<N<\xi(D_r)$.

An alternative way to extract the effective correlation length is
obtained by interpreting the abscissa of
the minimum of each curve in Fig. \ref{fig:E0_ND-E0_inf} as a length
proportional to  $\xi(D)$. Fitting a straight line through these minima
in a log-log plot of $N(D)$ yields for the IS model the exponent
$\kappa_{\textrm{IS}}\approx2.0293$ which approximates the analytical
result with an accuracy of roughly $0.24\%$.
On one hand this result is closer to the analytical value than the one
obtained using the degenerated hyperbola fit.
On the other hand, if we want to predict the bond dimension $D$ for which
the jump in the trace distance occurs in simulations with fixed $N$,
it turns out that the value obtained using this fit in does not always
coincide with the actual values observed in figure~\ref{fig:IS_TD_D64}.
As mentioned above, the degenerated hyperbola fit
satisfies this consistency test, which is why we prefer using that method
to extract an approximation for $\kappa$.
Furthermore the plots in figure~\ref{fig:E0_ND-E0_inf} require knowledge
of the exact ground state energy in the thermodynamic limit, which is not
always available. The strategy with the hyperbola fit on the other hand
does not require any analytical results and thus can always be used.

\section{Detailed treatment of the thermodynamic limit of the transition}
\label{sec:TL_of_transition}

In this appendix we present the details we used for the conclusion
drawn in section~\ref{sec:IS_TL} of the main text.
As mentioned above in appendix~\ref{sec:correction_to_scaling}
a reliable analysis of the thermodynamic
limit can only be made properly if we move towards it on paths of
constant $k=N/D^\kappa$. However this analysis provides conclusive
results only for the IS model which is why we skip presenting the
results obtained for the HB model. As mentioned in the main text,
the reason why this method fails for the HB model is that the
\emph{reference states} are in that case not precise enough.

As a first step let us normalize the tensors in our states such that the
largest eigenvalue of the MPS transfer matrix $T$ is equal to one (i.e.
$\lambda_1=1$ and $\lambda_i\geq\lambda_j,\,\,\,\forall i<j$). This
yields for the norm of such a state

\begin{equation}\label{eq:mps_norm_01}
  \braket{\Psi(D,N)|\Psi(D,N)}=\tr( T^N )
  = 1 + \sum_{i=2}^{D^2} \lambda_i^N(D,N)
  \,\,\,.
\end{equation}
\noindent
We will always use in the following lower-case greek letters to denote
states that are normalized to one and upper-case letters for the
corresponding state normalized according to (\ref{eq:mps_norm_01}), i.e.

\begin{equation}\label{eq:mps_norm_02}
  \ket{\psi}=\frac{\ket{\Psi}}{\sqrt{ \braket{\Psi|\Psi} }}
  \,\,\,.
\end{equation}
\noindent
For the computation of the trace distance between \emph{reference states}
and states lying on a curve with fixed $k$ we need the absolute square of
the overlap which becomes

\begin{equation}\label{eq:mps_overlap}
\begin{split}
  &\left| \braket{\psi(D_{k,N},N)|\psi(D_{\textrm{ref}},N)} \right|^2\\
  =&\frac{
          \left|
          \braket{\Psi(D_{k,N},N)|\Psi(D_{\textrm{ref}},N)}
          \right|^2
         }
         {
          \braket{\Psi(D_{k,N},N)|\Psi(D_{k,N},N)}
          \braket{\Psi(D_{\textrm{ref}},N)|\Psi(D_{\textrm{ref}},N)}
         }\\
  =&\frac{ \left[ \sum_{i=1}^{D_{\textrm{ref}}\cdot D_{k,N}}
                 \mu_i^N(D_{k,N},D_{\textrm{ref}},N) \right]^2}
  { \left[ 1 + \sum_{i=2}^{D^2_{k,N}} \lambda_i^N(D_{k,N},N) \right]
    \left[ 1 + \sum_{i=2}^{D_{\textrm{ref}}^2}
               \lambda_i^N(D_{\textrm{ref}},N) \right] }
  \,\,\,.
\end{split}
\end{equation}
\noindent
In the numerator we have used $\mu_i(D_{k,N},D_{\textrm{ref}},N) =: \mu_i(k,N)$
to denote the eigenvalues of the overlap transfer matrix

\begin{equation}\label{eq:TM_overlap}
  T_{\textrm{ovlp}}(k,N)
  = \sum_{i=1}^{d} A_i(D_{k,N},N) \otimes A_i^*(D_{\textrm{ref}},N)
  \,\,\,,
\end{equation}
\noindent
where the $A_i(D,N)$ represent as usually the matrices of a translationally
invariant MPS with $N$ sites and virtual bond dimension $D$.
Similarly we will use for the eigenvalues of the MPS transfer matrix
the notation $\lambda_i(k,N):=\lambda_i(D_{k,N},N)$ in the following.
This can be done since we need only two of the quantities $(D,N,k)$ to
uniquely specify the point of the phase diagram that we want to refer to.

The crucial argument in favor of the persistence of a discrete transition
between the two regimes in the thermodynamic limit will be the fact that
in this limit $\mu_1(k,N)$ converges much faster to $1$ in
the FSS regime (i.e. for $k<k_c$) than it does in the FES regime
(i.e. for $k>k_c$). In fact we will show below that in the first
case $\lim_{N\to\infty} \mu_1^N(k,N)=1$ while in the second case we have
$\lim_{N\to\infty} \mu_1^N(k,N)=0$. The other contributions in the
numerator of (\ref{eq:mps_overlap}) will turn out to be negligible for
$N\to\infty$, i.e. $\lim_{N\to\infty} \mu_i^N(k,N)=0$ for any $k$ and all
$i>1$. Furthermore we will show that the denominator of
(\ref{eq:mps_overlap}) remains finite in all cases such that we will be
able to conclude that the overlap of the \emph{quasi-exact}
\footnote{
  Quasi-exact means in this context that we use as a \emph{reference
  state} a MPS with virtual bond dimension $D$ that is \emph{much larger}
  than the one of the studied points on the path of constant $k$.
  If we restrict our scaling analysis to chains of length
  $N \ll \xi(D_{\textrm{ref}})$ it is sensible to assume that the
  MPS with bond dimension $D_{\textrm{ref}}$ is much closer to
  the exact ground state than it is to the states we analyze.
  Thus the overlap that we obtain in this way is very close to
  the overlap with the exact ground state.
}
ground state with states in the FES regime converges to zero in the
thermodynamic limit. Along the same lines we will argue that the overlap of
the \emph{quasi-exact} ground state with states in the FSES regime is
always larger than zero in the thermodynamic limit, thereby concluding
that a detectable transition between the two regimes persists for
$N\to\infty$.

To this end we have considered three paths in the FSS regime
($k\approx{0.37, 0.54, 0.97}$) and two paths in the FES regime
($k\approx{18.0, 58.7}$). The exact data points $(N,D)$ for four
of these paths are listed in table~\ref{tab:path_parameters}.
Note that since $N,D\in\mathbb{N}$ the exact value for
$k=N/D^\kappa$ varies slightly within each path.

\begin{table}[ht]
\begin{tabular}{|c|c|c||c|c|c||c|c|c||c|c|c|}
\hline
\multicolumn{12}{|c|}{$\kappa=2.03425$} \\
\hline
\multicolumn{3}{|c||}{$k\approx 0.37<k_c$} & \multicolumn{3}{|c||}{$k\approx 0.54<k_c$} & \multicolumn{3}{|c|}{$k\approx 18.0>k_c$} & \multicolumn{3}{|c|}{$k\approx 58.7>k_c$}\\
\hline
$k$     & $N$    & $D$  & $k$     & $N$    & $D$  & $k$    & $N$    & $D$ & $k$    & $N$    & $D$ \\ \hline
$0.374$ & $122$  & $17$ & $0.538$ & $118$  & $14$ & $17.9$ & $300$  & $4$ & $58.9$ & $1000$ & $4$\\
$0.373$ & $206$  & $22$ & $0.540$ & $198$  & $18$ & $17.8$ & $470$  & $5$ & $59.0$ & $1580$ & $5$\\
$0.371$ & $288$  & $26$ & $0.540$ & $298$  & $22$ & $18.3$ & $700$  & $6$ & $58.7$ & $2280$ & $6$\\
$0.371$ & $386$  & $30$ & $0.536$ & $384$  & $25$ & $18.1$ & $950$  & $7$ & $58.8$ & $3130$ & $7$\\
$0.369$ & $526$  & $35$ & $0.539$ & $560$  & $30$ & $18.2$ & $1250$ & $8$ & $58.6$ & $4100$ & $8$\\
$0.369$ & $690$  & $40$ & $0.537$ & $810$  & $36$ & $17.7$ & $1550$ & $9$ & $58.6$ & $5210$ & $9$\\
$0.368$ & $1000$ & $48$ & $0.535$ & $1000$ & $40$ & $18.0$ & $1950$ & $10$ & $58.5$ & $6450$ & $10$\\
        &        &      &         &        &      & $17.9$ & $2350$ & $11$ & $58.4$ & $7830$ & $11$\\
        &        &      &         &        &      & $17.9$ & $2800$ & $12$ & $58.4$ & $9350$ & $12$\\
\hline
\end{tabular}
\caption{Data points constituting several of the investigated paths with
roughly constant $k$ depicted in figure~\ref{fig:IS_TLpath_norm_TMeigval}.}
\label{tab:path_parameters}
\end{table}

\begin{figure}[ht]
  \begin{center}
    \includegraphics[width=1.0\columnwidth]{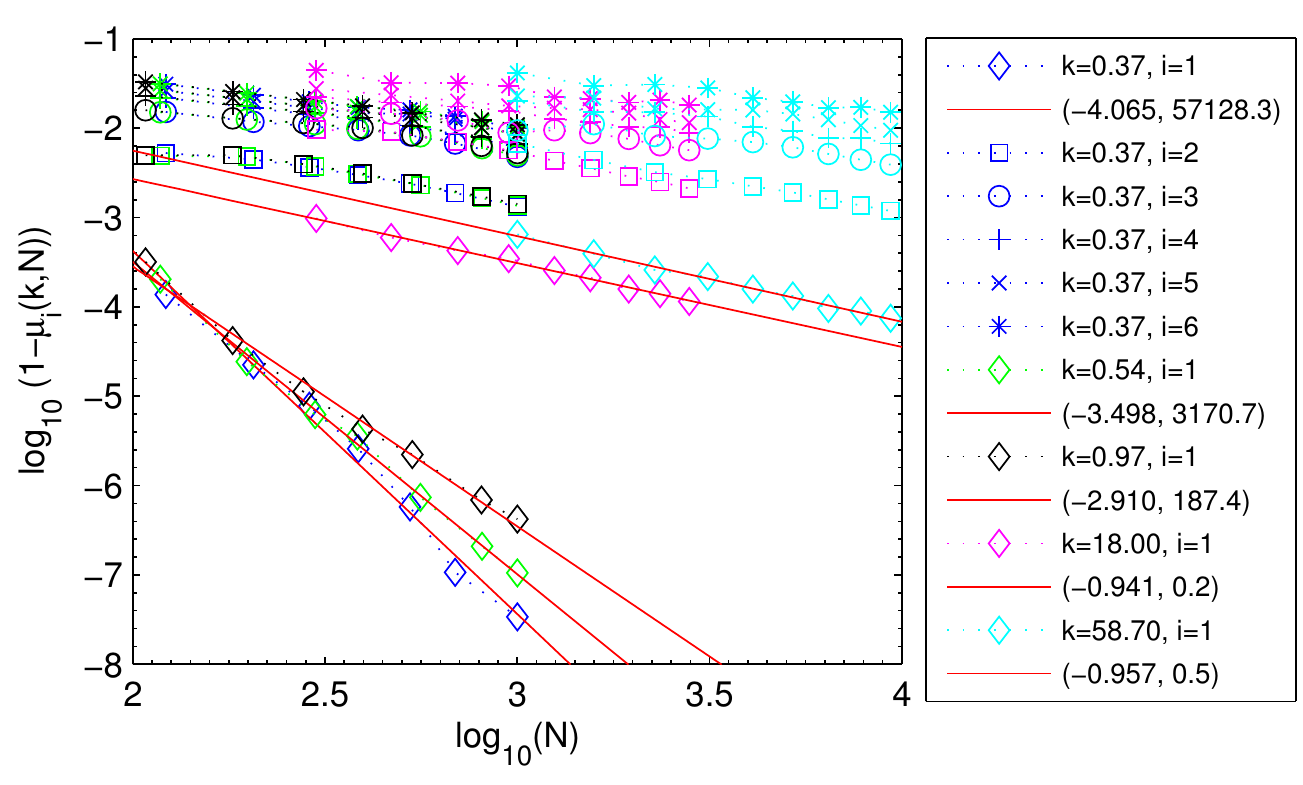}
  \end{center}
  \caption{
    (Color online).
    Quantum Ising model:
    scaling of the eigenvalues of the overlap transfer matrix for five
    paths with roughly constant $k=N/D^\kappa$ in different regimes.
    The exact pairs $(N,D)$ for the data points are given in
    table~\ref{tab:path_parameters}. In the legend we have only explained
    in detail what the different markers mean for the path with
    $k\approx 0.37$. The markers for the other paths follow the
    same pattern. The full red lines are linear fits through
    the data for each $\mu_1(k,N)$ respectively.
    The legend entries for these lines contain the values
    $(-\beta_1(k),\alpha_1(k))$.
  }
\label{fig:IS_TLpath_overlap_TMeigval}
\end{figure}

Let us first investigate how the numerator of the ratio
(\ref{eq:mps_overlap}) behaves.
We have observed that if we look at the eigenvalues
$\mu_i(k,N)$ along paths with constant $k$, then
$1-\mu_i(k,N)$ scales polynomially in $N$ as can be seen in the
log-log plot of figure~\ref{fig:IS_TLpath_overlap_TMeigval},
such that we have

\begin{equation}\label{eq:mui_scaling}
  \mu_i(k,N)=1-\frac{\alpha_i(k)}{N^{\beta_i(k)}}
  \,\,\,.
\end{equation}


\begin{table}
\begin{tabular}{|c||c|c||c|c||c|c|}
\hline
\multicolumn{7}{|c|}{$\kappa=2.03425$} \\
\hline
& \multicolumn{2}{|c||}{$k\approx 0.37<k_c$} & \multicolumn{2}{|c||}{$k\approx 0.54<k_c$} & \multicolumn{2}{|c|}{$k\approx 0.97<k_c$}\\
\hline
$i$  & $\beta_i$ & $\alpha_i$ & $\beta_i$ & $\alpha_i$ & $\beta_i$ & $\alpha_i$ \\ \hline
$1$  & $4.06477$ & $57128.2$ & $3.49773$ & $3170.6$  & $2.90998$ & $187.3$ \\
$2$  & $0.66454$ & $0.14622$ & $0.64177$ & $0.12814$ & $0.60463$ & $0.10416$ \\
$3$  & $0.51554$ & $0.19106$ & $0.51270$ & $0.19112$ & $0.48326$ & $0.16400$ \\
$4$  & $0.58048$ & $0.37486$ & $0.57835$ & $0.37607$ & $0.55326$ & $0.33117$ \\
$5$  & $0.48173$ & $0.26650$ & $0.51001$ & $0.31294$ & $0.51875$ & $0.32733$ \\
$6$  & $0.50660$ & $0.35546$ & $0.50828$ & $0.36216$ & $0.49502$ & $0.33813$ \\
$7$  & $0.47451$ & $0.31883$ & $0.46129$ & $0.30227$ & $0.42975$ & $0.26069$ \\
$8$  & $0.46673$ & $0.35474$ & $0.46928$ & $0.36415$ & $0.45436$ & $0.33884$ \\
$9$  & $0.48042$ & $0.42354$ & $0.47013$ & $0.41012$ & $0.44586$ & $0.37264$ \\
$10$ & $0.51091$ & $0.55418$ & $0.50601$ & $0.54819$ & $0.48431$ & $0.49597$ \\
\hline
\end{tabular}
\caption{Scaling of $\mu_i(k,N)$: parameters $\beta_i(k)$ and
$\alpha_i(k)$ for the fitting of $1-\mu_i(k,N)=\alpha_i(k)/N^{\beta_i(k)}$
for paths in the FSES regime.}
\label{tab:overlap_fit_FSES}
\end{table}


\begin{table}
\begin{tabular}{|c||c|c||c|c|}
\hline
\multicolumn{5}{|c|}{$\kappa=2.03425$} \\
\hline
& \multicolumn{2}{|c||}{$k\approx 18.0>k_c$} & \multicolumn{2}{|c|}{$k\approx 58.7>k_c$}\\
\hline
$i$  & $\beta_i$  & $\alpha_i$ & $\beta_i$  & $\alpha_i$ \\ \hline
$1$  & $0.94079$ & $0.20604$  & $0.95736$ & $0.46203$  \\
$2$  & $0.72904$ & $0.74874$  & $0.75036$ & $1.14015$  \\
$3$  & $0.46762$ & $0.25046$  & $0.43380$ & $0.23299$  \\
$4$  & $0.42154$ & $0.26209$  & $0.49195$ & $0.62481$  \\
$5$  & $0.37208$ & $0.23363$  & $0.37851$ & $0.33497$  \\
$6$  & $0.38713$ & $0.38566$  & $0.44232$ & $0.87896$  \\
$7$  & $0.36770$ & $0.38856$  & $0.38792$ & $0.65807$  \\
$8$  & $0.40633$ & $0.58072$  & $0.41378$ & $0.94025$  \\
$9$  & $0.42088$ & $0.80086$  & $0.45836$ & $1.67698$  \\
$10$ & $0.40215$ & $0.80529$  & $0.43306$ & $1.58778$  \\
\hline
\end{tabular}
\caption{Scaling of $\mu_i(k,N)$: parameters $\beta_i(k)$ and
$\alpha_i(k)$ for the fitting of $1-\mu_i(k,N)=\alpha_i(k)/N^{\beta_i(k)}$
for paths in the FESS regime.}
\label{tab:overlap_fit_FESS}
\end{table}


\noindent
Figure~\ref{fig:IS_TLpath_overlap_TMeigval} shows a log-log plot
of $1-\mu_i(k,N)$ for all $k$ and fixed $D_{ref}=64$.
The numerical values of $\alpha_i(k)$ and $\beta_i(k)$ for the $10$
largest $\mu_i(k,N)$ are listed for the paths in the FES regime in
table~\ref{tab:overlap_fit_FSES}. The equivalent data for
paths in the FES regime can be found in table~\ref{tab:overlap_fit_FESS}.
We see that in the FSS regime for $i=1$ we have $\beta_1(k)>1$
while in all other cases we get $\beta_i(k)<1$.
This means that for $N\to\infty$ the overlap (\ref{eq:mps_overlap})
always converges to zero in the FES regime due to

\begin{equation}\label{eq:explike_formula_zero}
  \lim_{N\to\infty} (1-\frac{\alpha}{N^{\beta}})^N=0
  \,\,\,\,\,\,\forall\beta<1,\alpha>0
\end{equation}
\noindent
and due to the fact that the denominator is always larger than zero
(in fact it is always larger than one).
In the FSS regime on the other hand, the $i=1$ terms in the numerator of
(\ref{eq:mps_overlap}) survive in the thermodynamic limit due to

\begin{equation}\label{eq:explike_formula_one}
  \lim_{N\to\infty} (1-\frac{\alpha}{N^{\beta}})^N=1
  \,\,\,\,\,\,\forall\beta>1,\alpha>0
  \,\,\,.
\end{equation}
\noindent
However this is not enough in order to show that the overlap is strictly
larger than zero in this regime. A diverging denominator in the limit
$N\to\infty$ could spoil this line of reasoning, so we have to convince
ourselves that both factors in the denominator of (\ref{eq:mps_overlap})
remain finite in the thermodynamic limit.

\begin{figure}[ht]
  \begin{center}
    \includegraphics[width=1.0\columnwidth]{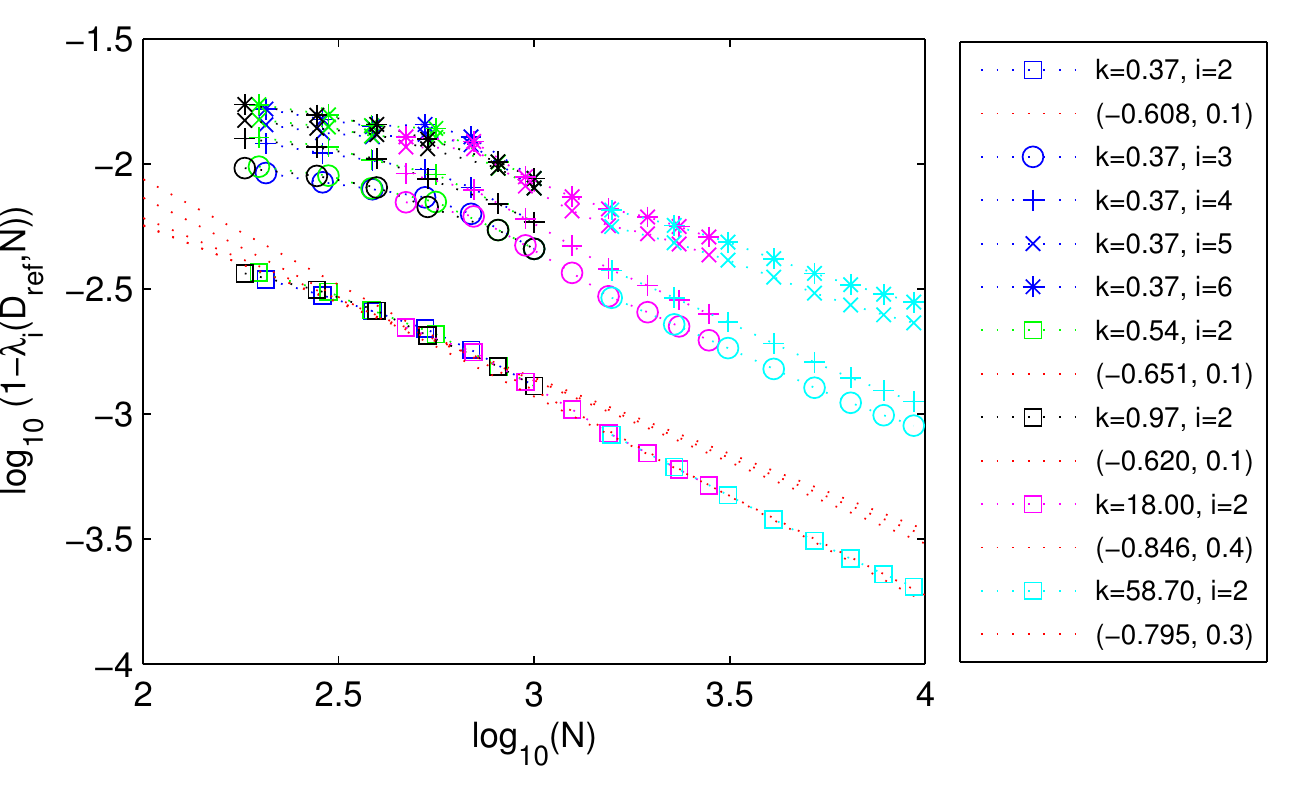}
  \end{center}
  \caption{
    (Color online).
    Quantum Ising model:
    scaling of the eigenvalues of the MPS transfer matrix for
    $D_{\textrm{ref}}=64$ and all $N$ occuring in
    table~\ref{tab:path_parameters}.
    In the legend we have only explained in detail what the
    different markers mean for the $N$ in the path with
    $k\approx 0.37$. The markers for the other paths follow the
    same pattern. The dotted red lines are linear fits through
    the data for each $\lambda_2(k,N)$ respectively.
    The legend entries for these lines contain the values
    $(-\beta_2(k),\alpha_2(k))$.
  }
\label{fig:IS_TLpathD64_norm_TMeigval}
\end{figure}


\begin{table}
\begin{tabular}{|c||c|c||c|c||c|c|}
\hline
\multicolumn{7}{|c|}{$\kappa=2.03425$} \\
\hline
& \multicolumn{2}{|c||}{$k\approx 0.37<k_c$} & \multicolumn{2}{|c||}{$k\approx 0.54<k_c$} & \multicolumn{2}{|c|}{$k\approx 0.97<k_c$}\\
\hline
$i$  & $\beta_i$ & $\alpha_i$  & $\beta_i$ & $\alpha_i$  & $\beta_i$ & $\alpha_i$ \\ \hline
$2$  & $0.60788$ & $0.09304$ & $0.65085$ & $0.12184$ & $0.62045$ & $0.09914$ \\
$3$  & $0.41227$ & $0.08847$ & $0.46692$ & $0.12433$ & $0.44144$ & $0.10401$ \\
$4$  & $0.43118$ & $0.12846$ & $0.48404$ & $0.17837$ & $0.45909$ & $0.14987$ \\
$5$  & $0.30225$ & $0.07647$ & $0.36915$ & $0.11385$ & $0.35609$ & $0.10226$ \\
$6$  & $0.34175$ & $0.10783$ & $0.40765$ & $0.15889$ & $0.39654$ & $0.14462$ \\
$7$  & $0.40046$ & $0.17074$ & $0.44681$ & $0.22606$ & $0.43958$ & $0.21126$ \\
$8$  & $0.41688$ & $0.19910$ & $0.46402$ & $0.26460$ & $0.45243$ & $0.24089$ \\
$9$  & $0.33412$ & $0.13003$ & $0.36584$ & $0.15862$ & $0.36051$ & $0.15116$ \\
$10$ & $0.28957$ & $0.10620$ & $0.34067$ & $0.14540$ & $0.33743$ & $0.14045$ \\
\hline
\end{tabular}
\caption{Scaling of $\lambda_i(D_{\textrm{ref}},k,N)$:
parameters $\beta_i(k)$ and $\alpha_i(k)$ for the fitting of
$1-\lambda_i(D_{\textrm{ref}},k,N)=\alpha_i(k)/N^{\beta_i(k)}$
for paths in the FSES regime.}
\label{tab:normDref_fit_FSES}
\end{table}


\begin{table}
\begin{tabular}{|c||c|c||c|c|}
\hline
\multicolumn{5}{|c|}{$\kappa=2.03425$} \\
\hline
& \multicolumn{2}{|c||}{$k\approx 18.0>k_c$} & \multicolumn{2}{|c|}{$k\approx 58.7>k_c$}\\
\hline
$i$  & $\beta_i$  & $\alpha_i$  & $\beta_i$  &  $\alpha_i$ \\ \hline
$2$  & $0.84555$ & $0.42728$ & $0.79472$ & $0.28481$  \\
$3$  & $0.76096$ & $0.82750$ & $0.67333$ & $0.41376$  \\
$4$  & $0.76823$ & $1.11696$ & $0.68922$ & $0.59758$  \\
$5$  & $0.61161$ & $0.54136$ & $0.51839$ & $0.26161$  \\
$6$  & $0.55554$ & $0.41101$ & $0.49502$ & $0.25624$  \\
$7$  & $0.42416$ & $0.19075$ & $0.52876$ & $0.42488$  \\
$8$  & $0.36840$ & $0.13854$ & $0.42626$ & $0.21306$  \\
$9$  & $0.42172$ & $0.22169$ & $0.40383$ & $0.19455$  \\
$10$ & $0.39138$ & $0.19807$ & $0.31502$ & $0.10918$  \\
\hline
\end{tabular}
\caption{Scaling of $\lambda_i(D_{\textrm{ref}},k,N)$:
parameters $\beta_i(k)$ and $\alpha_i(k)$ for the fitting of
$1-\lambda_i(D_{\textrm{ref}},k,N)=\alpha_i(k)/N^{\beta_i(k)}$
for paths in the FESS regime.}
\label{tab:normDref_fit_FESS}
\end{table}

Let us first treat the norm of the reference MPS
since this turns out to be the easier one.
Figure~\ref{fig:IS_TLpathD64_norm_TMeigval} shows a log-log plot
of $1-\lambda_i(k,N)$ for all $k$ and fixed $D_{ref}=64$.
The numerical values for $i\leq 10$ are given in
tables~\ref{tab:normDref_fit_FSES} and~\ref{tab:normDref_fit_FESS}.
For large chains with
$N>1000$ figure~\ref{fig:IS_TLpathD64_norm_TMeigval}
clearly indicates polynomial scaling in $N$.
Note that for small chains with $N<1000$ the plot deviates from
the nice linear behavior that we see for $N>1000$. The reason for this
are numerical errors in the computation of the ground state MPS.
This effect can also be seen in figure~\ref{fig:ISHB_Nall_Dall}:
for $N<1000$ the algorithm we use cannot minimze the energy
beyond a relative precision of roughly $8\cdot 10^{-11}$
even if we decrease $N$ while keeping a constant $D=64$.
Apart from that, the fitting in
figure~\ref{fig:IS_TLpathD64_norm_TMeigval} yields all
$\beta_i(k)<1$ for $i\geq 2$ thus we can conclude that
the norm $\braket{\Psi(D_{ref},N)|\Psi(D_{ref},N)}$
converges to one in the thermodynamic limit when we use the
normalization prescription (\ref{eq:mps_norm_01}).

\begin{figure}[ht]
  \begin{center}
    \includegraphics[width=1.0\columnwidth]{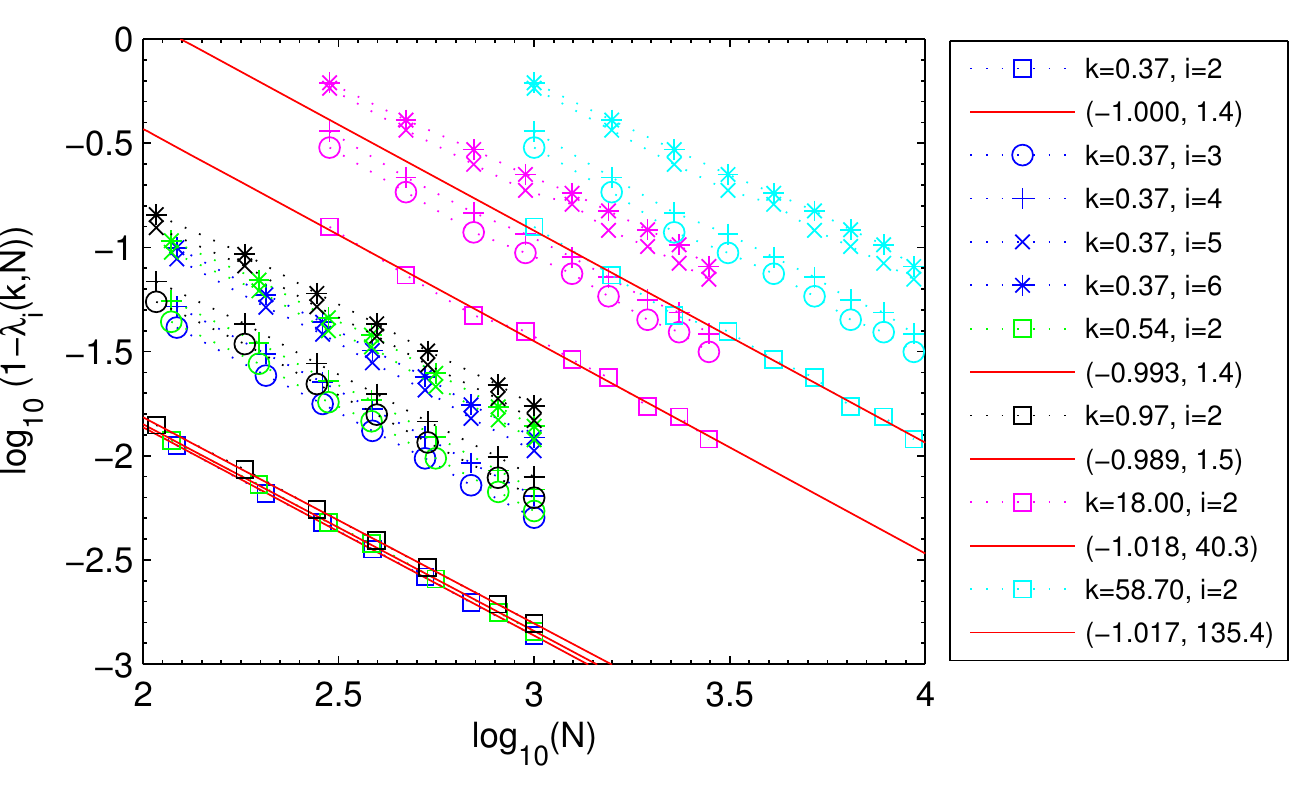}
  \end{center}
  \caption{
    (Color online).
    Quantum Ising model:
    scaling of the eigenvalues of the MPS transfer matrix for five
    paths with roughly constant $k=N/D^\kappa$ in different regimes.
    The exact pairs $(N,D)$ for the data points are given in
    table~\ref{tab:path_parameters}. In the legend we have only explained
    in detail what the different markers mean for the path with
    $k\approx 0.37$. The markers for the other paths follow the
    same pattern. The full red lines are linear fits through
    the data for each $\lambda_2(k,N)$ respectively.
    The legend entries for these lines contain the values
    $(-\beta_2(k),\alpha_2(k))$.
  }
\label{fig:IS_TLpath_norm_TMeigval}
\end{figure}


\begin{table}
\begin{tabular}{|c||c|c||c|c||c|c|}
\hline
\multicolumn{7}{|c|}{$\kappa=2.03425$} \\
\hline
& \multicolumn{2}{|c||}{$k\approx 0.37<k_c$} & \multicolumn{2}{|c||}{$k\approx 0.54<k_c$} & \multicolumn{2}{|c|}{$k\approx 0.97<k_c$}\\
\hline
$i$  & $\beta_i$ & $\alpha_i$  & $\beta_i$ & $\alpha_i$  & $\beta_i$ & $\alpha_i$ \\ \hline
$2$  & $0.99979$ & $1.36876$  & $0.99261$ & $1.37029$  & $0.98869$ & $1.45137$ \\
$3$  & $0.99838$ & $5.00085$  & $0.98501$ & $4.96160$  & $0.97751$ & $5.42433$ \\
$4$  & $0.99397$ & $6.19275$  & $0.98129$ & $6.13836$  & $0.97395$ & $6.67115$ \\
$5$  & $1.00992$ & $11.36417$ & $0.97979$ & $10.64645$ & $0.96439$ & $11.79694$ \\
$6$  & $0.99749$ & $12.14274$ & $0.96755$ & $11.35451$ & $0.95325$ & $12.79517$ \\
$7$  & $0.99290$ & $15.20975$ & $0.96478$ & $14.29878$ & $0.94933$ & $15.72285$ \\
$8$  & $0.99385$ & $16.84721$ & $0.94982$ & $14.52223$ & $0.93257$ & $16.11736$ \\
$9$  & $1.03853$ & $23.13857$ & $0.98133$ & $19.63216$ & $0.94593$ & $20.59742$ \\
$10$ & $1.03720$ & $25.64027$ & $0.95765$ & $19.24667$ & $0.92658$ & $21.15268$ \\
\hline
\end{tabular}
\caption{Scaling of $\lambda_i(k,N)$: parameters $\beta_i(k)$ and
$\alpha_i(k)$ for the fitting of $1-\lambda_i(k,N)=\alpha_i(k)/N^{\beta_i(k)}$
for paths in the FSES regime.}
\label{tab:norm_fit_FSES}
\end{table}


\begin{table}
\begin{tabular}{|c||c|c||c|c|}
\hline
\multicolumn{5}{|c|}{$\kappa=2.03425$} \\
\hline
& \multicolumn{2}{|c||}{$k\approx 18.0>k_c$} & \multicolumn{2}{|c|}{$k\approx 58.7>k_c$}\\
\hline
$i$  & $\beta_i$  & $\alpha_i$  & $\beta_i$  &  $\alpha_i$ \\ \hline
$2$  & $1.01831$ & $40.25148$  & $1.01731$ & $135.41657$  \\
$3$  & $0.98774$ & $81.55159$  & $0.98683$ & $264.69533$  \\
$4$  & $0.97128$ & $88.78991$  & $0.97053$ & $282.91558$  \\
$5$  & $0.92704$ & $111.61975$ & $0.92647$ & $337.75755$  \\
$6$  & $0.88126$ & $94.42375$  & $0.88080$ & $270.69784$  \\
$7$  & $0.88252$ & $104.24989$ & $0.88232$ & $299.97710$  \\
$8$  & $0.77072$ & $64.57797$  & $0.77072$ & $162.76623$  \\
$9$  & $0.79065$ & $82.78179$  & $0.78996$ & $212.48337$  \\
$10$ & $0.76184$ & $72.53475$  & $0.76185$ & $180.90762$  \\
\hline
\end{tabular}
\caption{Scaling of $\lambda_i(k,N)$: parameters $\beta_i(k)$ and
$\alpha_i(k)$ for the fitting of $1-\lambda_i(k,N)=\alpha_i(k)/N^{\beta_i(k)}$
for paths in the FESS regime.}
\label{tab:norm_fit_FESS}
\end{table}

The norm of the states $\ket{\Psi(D_{k,N},N)}$ along paths with constant
$k$ also turns out to converge to a finite value even though the argument
is a bit trickier in this case. The scaling of the largest eigenvalues
$\lambda_i(k,N)$ for each path is shown
in figure~\ref{fig:IS_TLpath_norm_TMeigval}. The numerical values
for $i\leq 10$ are given in tables~\ref{tab:norm_fit_FSES}
and~\ref{tab:norm_fit_FESS}. We see that most
of the $\beta_i(k)$ are very close to one for small $i$ in contrast
to the values obtained for $\lambda_i(D_{ref},N)$ which are all well
below one. In fact some of the $\beta_i(k)$ are even bigger than one
suggesting that $\lim_{N\to\infty}\lambda_i^N=1$ in these cases.
In section~\ref{sec:appendix_scaling_IS_TLpath_norm} of this appendix
we give evidence for the fact that even if $\beta_i(k)>1$ in some cases,
the number of these values remains finite for any $k$. Furthermore we argue
that in these cases it is reasonable to assume that we actually
have $\beta_i(k)=1$ which yields in the thermodynamic limit

\begin{equation}\label{eq:exp_formula}
  \lim_{N\to\infty} \lambda_i^N(k,N)=
  \lim_{N\to\infty} (1-\frac{\alpha_i}{N})^N=\exp(-\alpha_i)
  \,\,\,.
\end{equation}
\noindent
Summing up all relevant contributions then yields for the norm
of states in the different regimes

\begin{equation}\label{eq:norm_all_regimes}
\begin{split}
  \lim_{N\to\infty}
  \braket{\Psi(k,N)|\Psi(k,N)}_{k<k_c} &\approx 2.2 \\
  \lim_{N\to\infty}
  \braket{\Psi(k,N)|\Psi(k,N)}_{k>k_c} &\leq 2.0
  \,\,\,.
\end{split}
\end{equation}
\noindent
This allows us to approximate the overlap (with a \emph{quasi-exact} state)
towards which MPS simulations in different regimes converge to
(on the paths we considered) as

\begin{equation}\label{eq:overlap_all_regimes}
\begin{split}
  \lim_{N\to\infty}
  \braket{\psi(k,N)|\psi(D_{\textrm{ref}},N)}_{k<k_c} &\approx 0.45 \\
  \lim_{N\to\infty}
  \braket{\psi(k,N)|\psi(D_{\textrm{ref}},N)}_{k>k_c} &= 0
  \,\,\,.
\end{split}
\end{equation}
\noindent
Thus we can conclude that the thermodynamic limit of the overlap
in the FSS regime is always greater than zero proving that there
is indeed an discrete transition from the FSS regime to
the FES regime where the overlap becomes zero.

\subsection{Scaling of $\lambda_i(k,N)$}
\label{sec:appendix_scaling_IS_TLpath_norm}

The first ten parameters $\alpha_i$ and $\beta_i$ for the MPS transfer
matrix eigenvalues $\lambda_i(k,N)$
on paths in the FSES regime are given in table~\ref{tab:norm_fit_FSES},
the ones for paths in the FESS regime in table~\ref{tab:norm_fit_FESS}.
In the FESS regime we  have $\beta_2>1$ which then yields a
contribution of $\lim_{N\to\infty} \lambda_2^N(k)=1$ to the norm.
For $i>2$ we clearly see how the $\beta_i$ rapidly decay below
one, thereby making sure that the corresponding contributions to the norm
become zero in the thermodynamic limit. This means that if we
approach the thermodynamic limit on paths in the FESS regime and
always normalize the MPS according to (\ref{eq:mps_norm_01}),
i.e. $\lambda_1=1$, the norm of these states does not get bigger than two.
In fact it is very likely that the true contribution of $\lambda_2$ is
de facto zero
\footnote{
Actually it might also be that $\beta_2$ is in fact equal to one
which yields in the thermodynamic limit
$\lim_{N\to\infty}\lambda_2^N=\exp(-\alpha_2)$. Unfortunately,
as opposed to the similar case in the FSES regime, we cannot conclude
here that this must be the case.
}: for $N$ as big as $10^9$, using the values for
$\alpha_2$ and $\beta_2$ given in table~\ref{tab:norm_fit_FESS}, we get
$\lambda_2^N(k=18.0)\approx 2\cdot 10^{-12}$ and
$\lambda_2^N(k=58.7)\approx 4\cdot 10^{-42}$.

In the FSES regime on the other hand the $\beta_i$ seem to oscillate
randomly around one so we must look at the behavior of larger $i$ in
order to see if and when they decay below one,
which is what we ultimately need in order to show that the norm
of these states remains finite in the thermodynamic limit when the
normalization prescription (\ref{eq:mps_norm_01}) is employed.

\begin{figure}[ht]
  \begin{center}
    \includegraphics[width=1.0\columnwidth]{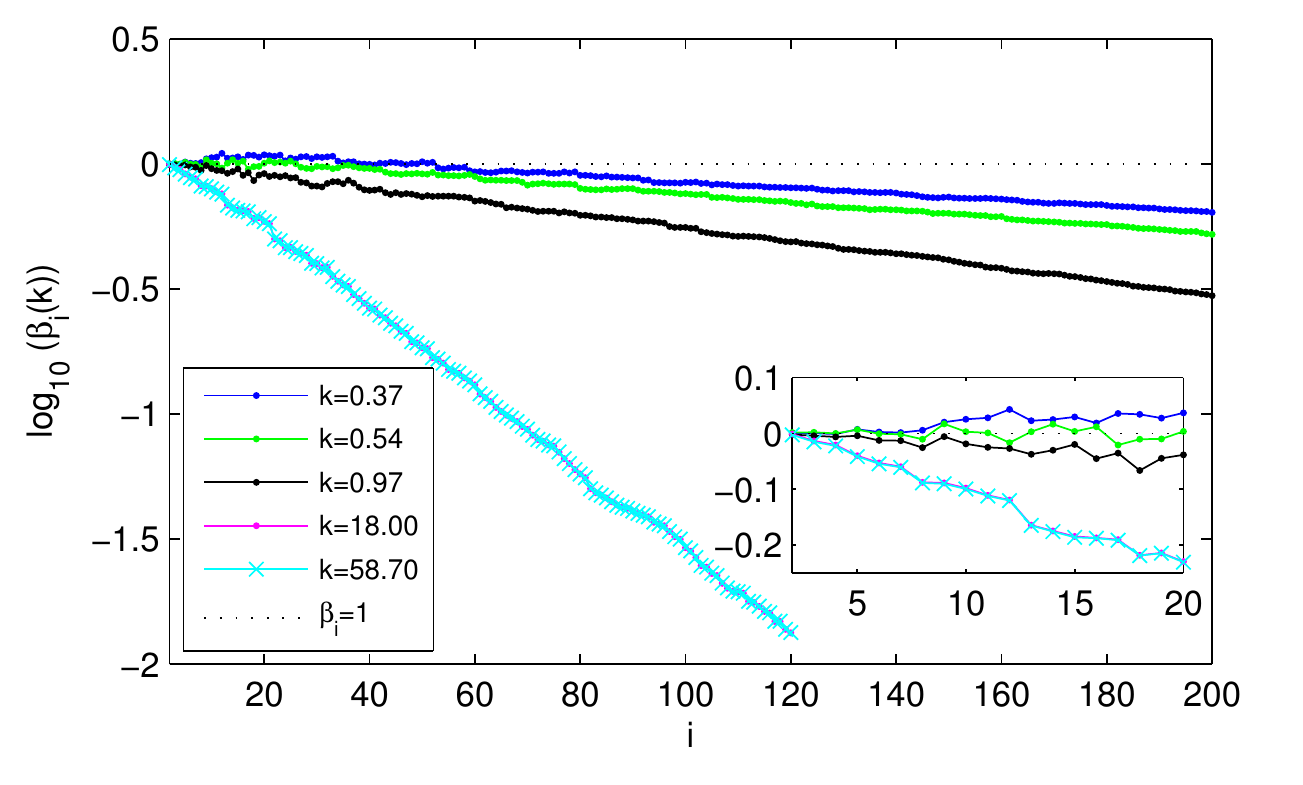}
  \end{center}
  \caption{
    (Color online).
    Quantum Ising model:
    log-plot of the first values of $\beta_i(k)$ for $i\leq 200$.
    A zoomed view on the first $20$ values is shown in the inset.
  }
\label{fig:IS_TLpath_norm_beta200}
\end{figure}

\begin{figure}[ht]
  \begin{center}
    \includegraphics[width=1.0\columnwidth]{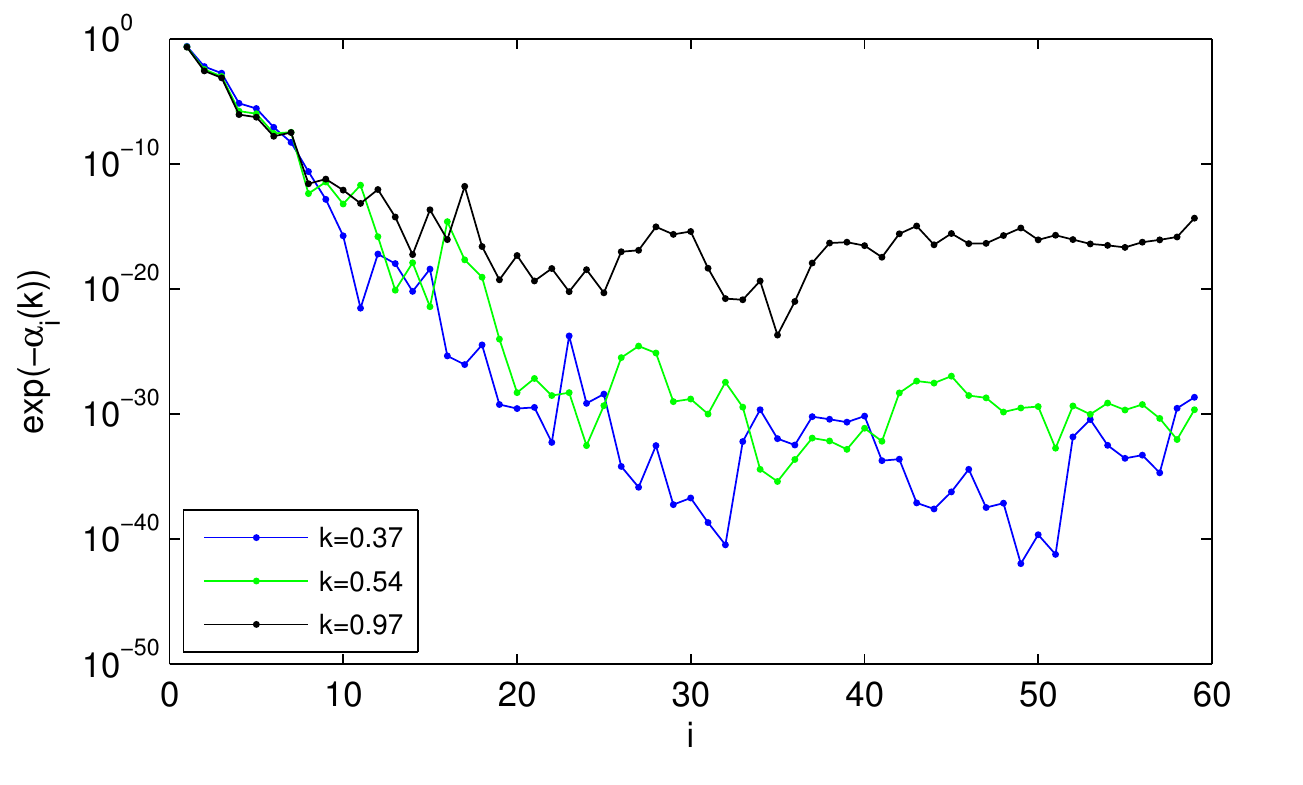}
  \end{center}
  \caption{
    (Color online).
    Quantum Ising model:
    contribution of the eigenvalues with $\beta_i(k)\approx 1$
    to the norm of the states in the thermodynamic limit,
    i.e. $\lim_{N\to\infty}\lambda_i^N(k)\approx\exp(-\alpha_i(k))$.
    Note how all contributions fall off exponentially
    below machine precision at $i\approx 14$.
  }
\label{fig:IS_TLpath_norm_alpha60}
\end{figure}

Figure~\ref{fig:IS_TLpath_norm_beta200} shows a log-plot of the first
$200$ $\beta_i$ in the FSES regime and of the first $120$ in the FESS
regime. All curves are approximately straight lines in this plot
which means that the $\beta_i$ decay exponentially with $i$.
Remember that on the paths we chose to investigate in the
FESS regime, the MPS with the largest virtual bond dimension have
$D=12$, thus we cannot fit any parameters $\beta_i$ for $i>121$
since we have only one data point available there. For $100<i\leq 121$
we have only two data points, namely the ones for $D=11$ and $D=12$
(see table~\ref{tab:path_parameters} in the main text) which is usually
not the best premise for an accurate fit. Nonetheless the $\beta_i$
fitted in this range obey the same exponential decay observed for smaller
$i$, where more data points are available.
The inset in figure~\ref{fig:IS_TLpath_norm_beta200} shows a
zoom into the region with $i\leq 20$. While for $i\leq 8$ all $\beta_i$
in the FSES regime are very close to one, we observe that for larger
$i$, the $k=0.37$ line is visibly above the $\beta_i=1$ line.
This would suggest that in the thermodynamic limit the eigenvalues
$\lambda_{i>8}$ would each yield a contribution equal to one to the
norm while the contribution from the $\lambda_i$ with
$i\in\{2,3,4,6,7,8\}$ would vanish, since in these cases $\beta_i<1$. This
makes however no sense since the $\lambda_i$ are decreasingly ordered,
i.e. $\lambda_i>\lambda_j$ if $i<j$. This leads us to the conclusion that
the oscillations around one that we observe for $i>8$ are numerical
relics and that the true value of the $\beta_i$ is either one or
something smaller than one. This conclusion is based on the fact that
in MPS simulations the transfer matrix eigenvalues that converge first
are the dominant eigenvalues (i.e. the ones with the largest absolute
value) so we can assume that the values obtained for $\beta_{i\leq 8}$
are much more accurate than the other ones.

Thus the worst case for our purpose is when all $\beta_i$ that are
not clearly smaller than one, are actually equal to one. Let us
investigate what we would get for the norm in this case.
If $\beta_i=1$, the contribution of these eigenvalues to the
norm in the thermodynamic limit solely depends on $\alpha_i$ due to

\begin{equation}\label{eq:exp_formula_02}
  \lim_{N\to\infty} \lambda_i^N(k,N)=1
  \lim_{N\to\infty} (1-\frac{\alpha_i}{N})^N=\exp(-\alpha_i)
  \,\,\,.
\end{equation}
\noindent
Figure~\ref{fig:IS_TLpath_norm_alpha60} shows the behavior of
$\exp(-\alpha_i)$ for the paths in the FSES regime and $i<60$,
which according to figure~\ref{fig:IS_TLpath_norm_beta200}
is the problematic $i$-range. We see how all contributions
rapidly decay below machine precision. Note that the black line
(i.e. the path with $k=0.97$) is for $i\geq 17$
several orders of magnitude above the $k=0.37$ and $k=0.54$ lines,
which is due to the fact that the corresponding $\beta_i$ are
so much smaller than one in this region, that the assumption
$\beta_i\approx 1$ simply does not hold, and the actual contribution
to the norm converges to zero. Note furthermore how for small $i$
all three lines are almost on top of each other meaning that the
values to which the norm converges in the thermodynamic limit
for MPS on different paths in the FSES regime will be very similar.
In fact we get

\begin{equation}\label{eq:norm}
\begin{split}
  \lim_{N\to\infty}
  \braket{\Psi(k,N)|\Psi(k,N)}_{k=0.37} &= 2.261646939734277 \\
  \lim_{N\to\infty}
  \braket{\Psi(k,N)|\Psi(k,N)}_{k=0.54} &= 2.236037631274709 \\
  \lim_{N\to\infty}
  \braket{\Psi(k,N)|\Psi(k,N)}_{k=0.97} &= 2.225635928039641 \\
\end{split}
\end{equation}
\noindent
which completes our argument that the norm of the MPS remains
finite on any path in the FSES regime.

\section{Comparison to other PBC MPS algorithms}
\label{sec:other_algorithms}

In this appendix we will show that the algorithm~\cite{me-2010-PBCI}
that we used to obtain all results in this work is performing
better than other recently presented approaches.

For the sake of completeness let us first recapitulate the
result of the comparison to the algorithm presented in~\cite{pippan-2010}.
We have already shown in~\cite{me-2010-PBCI} that our PBC algorithm yields
a better precision.
Apart from several other differences in these two approaches, the crucial
point is that we allow for a variable dimension $n$ of the dominant
subspace used to approximate big powers of the transfer matrix.
Even though this contributes a factor $n^2$ to the overall computational
cost $O(n^2 D^3)$, we have shown in~\cite{me-2010-PBCI} that there is
no way to get rid of the factor $n$ if one wants to reproduce the
correlation function throughout the entire PBC chain faithfully.
If the same $n$-scanning strategy would be employed in~\cite{pippan-2010},
probably the same precision level could be achieved, however
the computational cost in that algorithm would then scale like
$O(NnD^3)$. There is an additional factor $N$ in that scaling because
that approach is not translationally invariant.
The power of $n$ is reduced by one due to the fact that the
energy is minimized directly and not using the gradient.

Next we would like to compare our PBC algorithm to the one presented
in~\cite{zhou-2009}. In that work the authors simulate the critical
Quantum Ising Model by using Time Evolving
Block Decimation~\cite{vidal-2004} to locally update a translationally
invariant MPS which is then plugged into a chain with PBC geometry in
order to compute the energy. The weakness of that algorithm is that
the local update of the MPS tensors does not take into account the
boundary conditions: the fixed point MPS is exactly the same
like the one obtained when trying to approximate the ground state of an
infinite chain. In spite of this, the ground state energy can
be approximated quite well since the scaling of the computational
cost is only $O(nD^3)$ which allows the use of very large $D$.
Unfortunately there are no explicit plots of the precision of the ground
state energy in~\cite{zhou-2009} as a function of $D$.
From the abstract and footnote 4 of that work we deduce
that the simulation that yields the error $\approx 2.0\times 10^{-10}$
for the critical Quantum Ising PBC chain with $N=4800$ was done with
a MPS with bond dimension $D=200$.
We reach the same precision with
$D$ as small as $64$ as can be seen in figure~\ref{fig:ISHB_Nall_Dall}.
Due to the higher computational cost of our algorithm $D=200$ is out of
reach for us. Nonetheless we have computed an approximation of the ground
state of the infinite chain with a translationally invariant MPS with
$D=200$ (details of this are given below) and then plugged this MPS
into a PBC chain geometry with $N=4800$. The relative precision that
we obtained\emph{} using this strategy was
$\Delta_{rel}E_0(N=4800,D=200)\approx 1.39\cdot 10^{-10}$
which is in perfect
agreement with the claim made in~\cite{zhou-2009}.
However, if we take into account the fact that a PBC simulation
with $N=4800$ and $D=200$ is well in the FSES regime due to
$N\ll \xi(D=200)\approx 1.9 10^{5}$,
it becomes immediately clear that with $D=200$ one can
in principle reach a much better precision than $\approx 10^{-10}$.
In other words, the results obtained in~\cite{zhou-2009} correspond
to the cyan (light) lines in the left plot of
figure~\ref{fig:IS_Nall_D8D16_2linefit}.
While this is perfectly fine for simulations in the FESS regime,
if one is in the FSES regime, there is room for one or more
orders of magnitude of improvement of the relative precision.

There is another point worth mentioning regarding our PBC
algorithm~\cite{me-2010-PBCI}. In order to check the claims
made in~\cite{zhou-2009}, we needed to first approximate the ground
state of the infinite chain with an MPS with $D=200$.
For this we used a new method
called Time-Dependent Variational Principle~\cite{juthoTDVP-2011}.
We did this because TDVP converges much faster
than Imaginary Time Evolution based on Matrix Product
Operators~\cite{me-2010-MPOR} or iTEBD~\cite{vidal-2007}.
The relative precision we achieved with TDVP was
$\Delta_{rel}E_0(N=\infty,D=200)\approx 7.7\cdot 10^{-11}$.
We knew that this cannot be the best precision that can
be reached with $D=200$ since in~\cite{me-2010-MPOR} we get roughly
the same precision with $D$ as small as $128$.
Thus we ran the PBC algorithm for a huge chain with $N=10^{6}$
sites on top of the MPS obtained by TDVP.
Choosing as the parameters of that algorithm $m=1000$ and $n=100$
we managed to reduce the relative precision to $\approx 1.3 \cdot 10^{-11}$
which is in perfect agreement with the polynomial scaling
shown in figure 1 of~\cite{me-2010-MPOR}.
The lesson learned from this approach is that TDVP results
can be improved using our PBC algorithm well in the FESS regime.
We emphasize that running the PBC algorithm with small $n$
did not yield any improvement to the TDVP result. Only with
$n$ as large as $100$ we obtained the improved precision.
This is quite strange as when we compute the energy density
for the infinite chain, only one dominant eigenvector is used, i.e. $n=1$.
So it seems that even if additional dominant eigenvectors do not
enter the final computation of the ground state energy, they
do have an effect during the local optimization procedure of the
translationally invariant MPS.


\end{appendix}


\end{document}